\documentclass[12pt, a4paper]{article}

\usepackage[utf8]{inputenc}
\usepackage{amsfonts,amsmath,amssymb}
\usepackage{bbold}
\usepackage{dsfont}
\usepackage{multicol}
\usepackage{cite}
\usepackage{slashed}
\usepackage{tikz}
\usepackage[compat=1.1.0]{tikz-feynman}
\usepackage{bigints}
\usepackage{physics}

\usepackage[
   top=10mm,
   bottom=20mm,
   left=20mm,
   right=20mm]{geometry}

\usepackage{xcolor}
\usepackage{authblk,ulem}
\usepackage[hidelinks]{hyperref}
\usepackage{caption}
\usepackage{graphicx}

\newcommand{\nn}{\nonumber \\}
\newcommand{\beq}{\begin{equation}}
\newcommand{\eeq}{\end{equation}}
\newcommand{\ba}{\begin{array}}
\newcommand{\ea}{\end{array}}
\newcommand{\bea}{\begin{eqnarray}}
\newcommand{\eea}{\end{eqnarray} }
\newcommand{\bal}{\begin{align}}
\newcommand{\eal}{\end{align}}

\newcommand{\ii}{\mathrm{i}}

\newcommand{\Ree}{\operatorname{Re}}
\newcommand{\M}{\mathcal{M}}
\newcommand{\GammaV}{\Gamma}
\newcommand{\PS}{\mathrm{d}\Phi_4}

\usepackage{slashed}
\usepackage{tikz}
\usepackage{tikz-feynman}
\tikzfeynmanset{compat=1.1.0}

\usepackage{subcaption}
\usetikzlibrary{calc}
\tikzfeynmanset{
  crossed dot/.style={
    /tikz/circle,
    /tikz/draw,
    /tikz/minimum size=7pt,
    /tikz/inner sep=0pt,
    /tikz/path picture={
      \draw
        (path picture bounding box.south west) --
        (path picture bounding box.north east);
      \draw
        (path picture bounding box.north west) --
        (path picture bounding box.south east);
    }
  }
}

\title{NNLO Final-State Quark-Pair Corrections in Implicit Regularization}

\author[1]{Ricardo J. C. Rosado}

\author[2]{Adriano Cherchiglia}

\author[3]{Marcos Sampaio}

\author[1]{Brigitte Hiller}

\affil[1]{CFisUC, Department of Physics, University of Coimbra, P-3004-516 Coimbra,
Portugal}
\affil[2]{Centro de Ciências Naturais e Humanas, Universidade Federal do ABC, 09210-580 , Santo Andr\'e, Brasil}
\affil[3]{Programa de P\'os-Gradua\c{c}\~ao em Modelagem Matem\'atica e Computacional (PPGMMC)\\
Centro Federal de Educa\c{c}\~ao Tecnol\'ogica de Minas Gerais (CEFET-MG), Belo Horizonte, MG, Brasil}

\begin{document}

\maketitle

\begin{abstract}
We investigate the NNLO final-state quark-pair correction to the decay of a neutral axial-vector current ($Z$) into massless quarks within Implicit Regularization (IReg), in the limit of large $N_f$. Our calculation tests, at two loops, the  recursive subtraction of ultraviolet subdivergences and the cancellation of infrared singularities between virtual and real contributions in a framework that does not rely on analytical continuation in spacetime dimension. After renormalization and phase-space integration, the regulator-dependent logarithms cancel and the known $N_f$ contribution to the inclusive rate is recovered. The result provides a non-trivial NNLO check of gauge invariance, renormalization consistency, and perturbative unitarity in IReg. In precision multi-loop calculations, fixed-dimension regularization methods can be advantageous by avoiding the dimensional continuation of intrinsically four-dimensional structures, such as $\gamma_5$ matrices and Levi-Civita tensors, thereby reducing ambiguities and symmetry-restoring counterterms.
\end{abstract}

\section{Introduction}

Precision tests of the Standard Model increasingly rely on the evaluation of radiative corrections beyond next-to-leading order. This need is particularly acute in the precision programs of the HL-LHC and future lepton colliders, where theoretical uncertainties must be controlled at the same level as the anticipated experimental accuracy \cite{ZurbanoFernandez:2020cco,FCC:2018evy,Heinemeyer:2021rgq}. In quantum chromodynamics, next-to-next-to-leading-order calculations are complicated not only by the growing number of loop and phase-space integrals, but also by the simultaneous occurrence of ultraviolet (UV) and infrared (IR) soft and collinear singularities.

At NNLO, a generic decay rate receives double-virtual contributions, consisting of the interference of the two-loop amplitude with the Born amplitude together with the square of the one-loop amplitude; real--virtual contributions, obtained from the interference of one-loop and tree-level amplitudes with one additional final-state parton; and double-real contributions, arising from tree-level amplitudes with two additional final-state partons. After ultraviolet renormalization, these contributions remain separately infrared divergent, but their sum is finite for an infrared-safe inclusive decay rate, in which all relevant partonic final states are summed over and their complete phase spaces, including the soft and collinear regions, are integrated \cite{Kinoshita1962,LeeNauenberg1964}. The organization of this cancellation at NNLO is one of the central problems addressed by modern subtraction and four-dimensional formulations \cite{TorresBobadilla:2020ekr}. A viable regularization prescription must therefore preserve the relations among contributions with different final-state multiplicities while maintaining gauge invariance, locality, and unitarity.

Dimensional regularization is the standard framework for higher-order calculations in gauge theories \cite{tHooftVeltman1972}, while comparisons among commonly used dimensional and four-dimensional regularization schemes can be found in Refs.~\cite{Gnendiger:2017pys,Bruque:2018bmy}. Its efficiency is connected with its compatibility with gauge symmetry and with the factorized structure of infrared singularities \cite{Catani1998}. Nevertheless, continuation away from the physical dimension introduces a number of technical and conceptual complications. Because the Lorentz and Dirac algebras must be extended to $d$ dimensions, evanescent structures may be generated. This is particularly delicate in chiral theories. Dimensional regularization is especially subtle in the presence of the $\gamma_5$ Dirac matrix, since its intrinsically four-dimensional definition cannot be extended to $d\neq4$ while simultaneously preserving its anticommutation properties, trace identities, and the relevant chiral symmetries  \cite{breitenlohner1977dimensionally,Jones:1982zf,korner1992practicable,larin1993renormalization,kreimer1994role,jegerlehner2001facts,belusca2021two}. Recent work has made this issue increasingly systematic: finite symmetry-restoring counterterms have been constructed for general chiral gauge theories, the role of evanescent structures has been clarified, and explicit Breitenlohner - Maison -'t Hooft - Veltman  (BMHV)  \cite{ breitenlohner1977dimensionally, Breitenlohner:1977hr} renormalization has been carried out through three and four loops in an Abelian chiral theory and through two loops in a non-Abelian chiral theory \cite{Cornella:2022hkc,Belusca-Maito:2023wah,Stockinger:2023ndm,Ebert:2024xpy,OlgosoRuiz:2024dzq,Kuhler:2025znv,vonManteuffel:2025swv}.

These issues have motivated the development of approaches in which the algebra is performed directly in four dimensions. In this context, higher-order calculations provide essential tests, since prescriptions that are equivalent at one loop need not remain equivalent once nested UV divergences and non-trivial real--virtual cancellations are present.

Related dimensional schemes, such as dimensional reduction and the four-dimensional-helicity scheme, retain parts of the algebra in a quasi-four-dimensional space but require the consistent treatment of evanescent degrees of freedom beyond one loop \cite{stockinger2005regularization,signer2009using,kilgore:2012tb}. Fully four-dimensional strategies include four-dimensional unsubtraction based on loop--tree duality \cite{hernandez2016towards,sborlini2016four}, Four-Dimensional Regularization/Renormalization \cite{Pittau:2012zd,Pittau:2018frs}, and Implicit Regularization \cite{BattistelMotaNemes1998,SampaioEtAl2006,CherchigliaSampaioNemes2011}. These approaches differ in how ultraviolet subtractions and real--virtual infrared cancellations are organized, but share the goal of retaining the physical-dimensional algebra for observable quantities \cite{Gnendiger:2017pys,TorresBobadilla:2020ekr}. Recent causal loop--tree duality developments extend this program to local multiloop UV renormalization and to causal-unitary constructions in which decay rates are generated from vacuum amplitudes and real--virtual singularities cancel locally \cite{Rios-Sanchez:2024xtv,ramirez2024rewording,LTD:2024yrb}.

Implicit Regularization (IReg) is a momentum-space framework formulated without changing the physical dimension of the theory \cite{BattistelMotaNemes1998,SampaioEtAl2006,CherchigliaSampaioNemes2011}. In IReg, the UV-divergent content of an amplitude is expressed in terms of basic divergent integrals that are independent of the external momenta. The finite momentum-dependent part is safely and unambiguously evaluated, while differences between divergent tensor and divergent scalar integrals with appropriate metric tensor combinations as factors are parametrized by surface terms. In the constrained version of the method, these surface terms are fixed by momentum-routing invariance, which is sufficient to preserve the Ward and Slavnov--Taylor identities in anomaly-free amplitudes \cite{FerreiraEtAl2012,SampaioEtAl2006}. The method can be organized recursively at arbitrary loop order and reproduces the Bogoliubov--Parasiuk--Hepp--Zimmermann (BPHZ) \cite{Bogoliubov:1957pq,Hepp:1966eg,Zimmermann:1969jw} subtraction structure, thereby preserving locality and the identity of divergent subgraphs when they are embedded into higher-loop diagrams \cite{CherchigliaSampaioNemes2011,AriasPerdomoEtAl2021}.

Applications of IReg to gauge theories have established its consistency at two loops, including the evaluation of renormalization-group functions in abelian and non-abelian models \cite{DiasEtAl2008,Cherchiglia:2020iug,Cherchiglia:2021yxz}, as well as in a supersymmetric setup \cite{Fargnoli:2010mf,Cherchiglia:2015vaa}.  Other applications and comparisons include the decays $h\to\gamma\gamma$ and $H\to gg$, the analysis of infrared cancellation without evanescent fields, and explicit connections between IReg and dimensional schemes \cite{Cherchiglia:2012zp,Gnendiger:2017pys,Pereira:2022mne,Cherchiglia:2020iug}. More recently, quark--antiquark decays mediated by vector, axial-vector, scalar, and pseudoscalar currents were studied at NLO \cite{Rosado:2023ist}. These calculations illustrate an important practical advantage of IReg for chiral processes: ultraviolet renormalization and the cancellation of infrared singularities can be implemented while the Dirac algebra and loop integrations are kept in the physical dimension, thereby avoiding many of the complications associated with a fully $d$-dimensional treatment of $\gamma_5$.

Although the formulation of IReg in the physical dimension does not make the treatment of $\gamma_5$ entirely automatic, it still offers important practical advantages for chiral calculations. For many NLO amplitudes in which $\gamma_5$ appears only at interaction vertices, consistent results can be obtained by symmetrizing closed fermion traces or by adopting a fixed ordering prescription for open fermion strings. The relation between momentum-routing invariance, fixed-dimensional Dirac algebra, and chiral anomalies has been analyzed in Refs.~\cite{Viglioni:2016nqc,Bruque:2018bmy,Cherchiglia:2021uce}. More general amplitudes, particularly those involving $\gamma_5$ inside fermion propagators \cite{Rosado:2024pyy} or at higher loop order, may require the quasi-four-dimensional extension
\begin{equation}
Q4S=4S\oplus X,
\end{equation}
in which the evanescent sector $X$ keeps track of finite remnants generated by the extended Dirac algebra and permits the correct anomaly structure to be recovered. This extension is, however, essentially algebraic: loop integrations remain in the physical dimension, while the separation of ultraviolet divergences into basic divergent integrals and the BPHZ-like renormalization structure of IReg are preserved. The $\gamma_5$ caveat therefore limits the naive use of strictly four-dimensional identities within divergent amplitudes, but does not remove the usefulness of IReg as a systematic framework for chiral theories \cite{Cherchiglia:2026lps}.

In the NLO applications, the inclusive rates were shown to satisfy the KLN theorem without introducing evanescent fields or additional symmetry-restoring counterterms \cite{Rosado:2023ist}. Extending the analysis to NNLO nevertheless represents a qualitatively more demanding test of the method. At this order, ultraviolet subdivergences, higher-rank tensor integrals, and infrared-singular virtual and real contributions must be treated consistently, while preserving the correspondence between a loop subgraph and the real-emission process obtained from its unitarity cut.

The inclusive process
\begin{equation}
    Z\longrightarrow q\bar q+X,
\end{equation}
where $X$ denotes additional unresolved partonic radiation, provides a particularly suitable benchmark for this purpose. Its lowest-multiplicity partonic channel is $Z\to q\bar q$, while higher-order corrections include final states with additional gluons or quark--antiquark pairs. The corresponding inclusive rate is directly related to the hadronic $R$ ratio in electron--positron annihilation and is known independently in perturbative QCD \cite{DineSapirstein1979,Page:2018ljf}. The process therefore offers a clean setting in which to test the regularization and renormalization framework, rather than to determine a previously unknown observable. At the same time, it is sufficiently non-trivial to probe current conservation, the tensor structure of the quark form factor, the consistent subtraction of ultraviolet subdivergences, and the cancellation of infrared singularities between virtual and real contributions.

In the present work we concentrate on the fermionic contribution proportional to $N_f$ at second order on the $\alpha_s$. In the virtual sector, this contribution contains a quark vacuum-polarization insertion embedded in the gluon line of the two-loop vertex correction. Its corresponding unitarity cut produces an additional unresolved quark--antiquark pair in the final state. The close relation between the virtual fermion bubble and its real cut makes this sector simpler than the complete NNLO correction, but it does not make it trivial. Rather, it isolates several of the structural requirements that a four-dimensional regularization prescription must satisfy.

At NNLO, a generic decay rate is decomposed into double-virtual, real--virtual, and double-real contributions. After ultraviolet renormalization, these terms are generally still separately infrared divergent, whereas their sum is finite for an infrared-safe inclusive observable, in which all relevant partonic final states and their soft and collinear regions are included.

In the fermionic contribution considered here, the virtual correction contains a UV-divergent fermion-bubble subgraph, while its unitarity cut produces the corresponding four-particle real-emission channel. The calculation therefore provides a simultaneous test of transversality, the consistent treatment of subdivergences and overall divergences, and the cancellation of soft and collinear singularities between virtual and real contributions. Using the same auxiliary mass scale $\mu$ in both sectors makes the cancellation of regulator-dependent logarithms explicit after renormalization and phase-space integration. Since momenta, polarization states, and Dirac algebra remain four-dimensional, the result also tests whether IReg preserves the correspondence between an infrared-singular loop subgraph and its on-shell cut at NNLO.

A closely related calculation was performed within Four-Dimensional Regularization/Renormalization (FDR), where the NNLO final-state quark-pair corrections to the inclusive $H\to b\bar b+X$ and $\gamma^*\to q\bar q+X$ decay rates\footnote{In this case only a vector was considered, not an axial-vector.} were evaluated in the large-$N_f$ limit of QCD, with the virtual and real contributions combined without recourse to dimensional regularization \cite{Donati:2013voa,Page:2018ljf}. That analysis demonstrated that this class of observables provides a sensitive test of gauge invariance, subintegration consistency, and unitarity in a four-dimensional framework. The present calculation addresses the same general problem from the conceptually distinct viewpoint of IReg, in which UV divergences are retained explicitly as basic divergent integrals and removed through an ordinary local renormalization procedure consistent with BPHZ.

Our purpose is thus twofold. We first develop the IReg treatment of the axial-vector two-loop vertex containing the fermion-bubble insertion, paying particular attention to transversality, momentum-routing invariance, and the separation of subdivergent and overall UV contributions. We then evaluate the corresponding real quark-pair contribution and combine it with the renormalized virtual result. The cancellation of the infrared divergences parametrized in terms of an auxiliary-mass and the recovery of the known inclusive coefficient provide a non-trivial NNLO test of IReg in massless QCD.

The paper is organized as follows. In Sec.~2 we summarize the IReg rules required for multiloop amplitudes and discuss the treatment of UV and infrared singularities. In Sec.~3 we apply the method to the fermionic $N_f$ contribution to a neutral-current decay into a massless quark--antiquark pair, with the detailed calculation written for $Z\to q\bar q$ and the purely vector limit corresponding to the $\gamma^*$ studied in \cite{Page:2018ljf}. We first analyze the virtual two-loop amplitude and its UV renormalization, and then calculate the real four-particle contribution. The cancellation of the infrared regulator and the final inclusive result are presented before our conclusions. We also provide some appendixes with complementary results.

\section{Implicit Regularization: Basic Rules}

%%%%%%%%%%%%%

Implicit Regularization (IReg) is a momentum-space framework in which the
ultraviolet-divergent content of a Feynman amplitude is separated algebraically
from its finite, momentum-dependent part, without continuing the loop
integrations away from the physical dimension
\cite{BattistelMotaNemes1998,SampaioEtAl2006,CherchigliaSampaioNemes2011}. The divergent part is expressed in terms of
external momentum-independent objects, called basic divergent integrals (BDIs), whereas
the finite part can be unambiguously evaluated. The method does not require the BDIs to be explicitly evaluated;
they are absorbed into local renormalization constants in the usual way.

The implementation of IReg involves three conceptually distinct operations:
first, the integrand must be written in a consistent normal form; second, its
ultraviolet-divergent part must be separated from the finite remainder; and
third, tensor BDIs must be reduced to scalar ones in a manner compatible with
momentum-routing invariance. At multiloop order, these operations are applied
recursively so that ultraviolet subdivergences and the overall divergence are
identified separately, reproducing the structure of the BPHZ subtraction
algorithm.

\subsection{Normal form of the integrand}

Before the IReg decomposition is applied, the numerator of the amplitude must
be brought to a normal form. Internal-symmetry algebra, Lorentz contractions,
and the admissible Dirac-algebra manipulations are carried out first.
Numerator--denominator consistency is then imposed: whenever a factor of the
squared loop momentum in the numerator coincides with a denominator factor, it
must be cancelled algebraically before regularization. For example,
\begin{equation}
    \left.
    \int_k
    \frac{k^2}
         {k^2 q^2 (k-q)^2}
    \right|_{\mathrm{IReg}}
    =
    \left.
    \int_k
    \frac{1}
         {q^2 (k-q)^2}
    \right|_{\mathrm{IReg}},
    \qquad
    \int_k\equiv\int\frac{d^4k}{(2\pi)^4}.
    \label{eq:numerator-denominator-consistency}
\end{equation}
This condition ensures that algebraically identical integrands are assigned
the same regularized expression. Likewise symmetric integration cannot be imposed
inside divergent integrals. In particular,
\begin{equation}
   \left[
   \int_k k^\mu k^\nu f(k^2)
   \right]_{\mathrm{IReg}}
   \neq
   \frac{g^{\mu\nu}}{4}
   \left[
   \int_k k^2 f(k^2)
   \right]_{\mathrm{IReg}}
   \label{eq:no-symmetric-integration}
\end{equation}
between divergent quantities. 

Dimension-specific identities must also be used with care. In particular,
the naive anticommutation of $\gamma_{5}$ through a Dirac string may change
the ordering of the matrices and, consequently, the tensor structure of the
loop momentum appearing in the numerator. For convergent integrals such
rearrangements are harmless, but in ultraviolet-divergent amplitudes they may
differ by finite terms, because Lorentz contractions and
regularization do not in general commute. The Dirac and Lorentz algebra must
therefore be reduced according to a fixed prescription before the IReg rules
are applied. Once this normal form is established, IReg acts linearly, leaves
power-counting-finite terms unchanged, and separates the amplitude uniquely
into finite and ultraviolet-divergent parts. A more detailed discussion of the above points can be found in \cite{Cherchiglia:2020iug}.

\subsection{Algebraic separation of the ultraviolet behavior}

Consider a massless propagator depending on an external momentum $p_i$. A
fictitious mass $\mu$ is introduced according to
\begin{equation}
    \frac{1}{(k-p_i)^2}
    \longrightarrow
    \frac{1}{(k-p_i)^2-\mu^2}.
\end{equation}
The following exact algebraic identity is then applied:
\begin{align}
    \frac{1}{(k-p_i)^2-\mu^2}
    &=
    \sum_{j=0}^{N_i-1}
    \frac{(-1)^j
    \left(p_i^2-2p_i\cdot k\right)^j}
    {(k^2-\mu^2)^{j+1}}
    \nonumber\\
    &\quad+
    \frac{(-1)^{N_i}
    \left(p_i^2-2p_i\cdot k\right)^{N_i}}
    {(k^2-\mu^2)^{N_i}
    \left[(k-p_i)^2-\mu^2\right]} .
    \label{eq:ireg-propagator-identity}
\end{align}
The integer $N_i$ is chosen as the smallest value for which the last term is
ultraviolet finite in the loop momentum $k$. We emphasize that equation
\eqref{eq:ireg-propagator-identity} is not a truncated Taylor expansion since the
last term retains the complete dependence on the external momentum, making
the identity exact.

The finite sum in the first line isolates the ultraviolet asymptotic
behavior. Its denominators depend only on the loop momentum and on the
auxiliary mass $\mu$, whereas all dependence on the physical external
momentum appears polynomially in the numerator. The last term is
power-counting finite and is therefore evaluated without modification.

\subsection{Basic divergent integrals and surface terms}

In four dimensions, the one-loop scalar logarithmic and quadratic BDIs are
defined by
\begin{equation}
    I_{\log}(\mu^2)
    \equiv
    \int_k\frac{1}{(k^2-\mu^2)^2},
    \qquad
    I_{\mathrm{quad}}(\mu^2)
    \equiv
    \int_k\frac{1}{k^2-\mu^2}.
    \label{eq:scalar-bdis}
\end{equation}
Tensor BDIs are similarly defined as
\begin{align}
    I_{\log}^{\mu_1\cdots\mu_{2r}}(\mu^2)
    &\equiv
    \int_k
    \frac{
    k^{\mu_1}\cdots k^{\mu_{2r}}
    }
    {(k^2-\mu^2)^{r+2}},
    \label{eq:tensor-log-bdi}
    \\
    I_{\mathrm{quad}}^{\mu_1\cdots\mu_{2r}}(\mu^2)
    &\equiv
    \int_k
    \frac{
    k^{\mu_1}\cdots k^{\mu_{2r}}
    }
    {(k^2-\mu^2)^{r+1}}.
    \label{eq:tensor-quad-bdi}
\end{align}
These objects contain no physical masses or external momenta. Thus, all
kinematic information remains in the finite part, while the ultraviolet
content is encoded in a universal set of BDIs. The regulator is not explicitly introduced,
and no numerical value is assigned to a BDI. Only relations among BDIs that
follow from consistency and symmetry requirements are imposed.

Tensor BDIs are reduced to scalar ones by separating total derivatives in
momentum space. For example,
\begin{align}
    \int_k
    \frac{\partial}{\partial k_\mu}
    \left(
        \frac{k^\nu}{(k^2-\mu^2)^2}
    \right)
    &=
    g^{\mu\nu}I_{\log}(\mu^2)
    -
    4I_{\log}^{\mu\nu}(\mu^2)
    \nonumber\\
    &\equiv
    \Upsilon^{\mu\nu}.
    \label{eq:surface-term}
\end{align}
The quantity $\Upsilon^{\mu\nu}$ is a surface term. It is the difference
between two divergent integrals with the same superficial degree of
divergence and is therefore finite but undetermined until a consistency
condition is imposed. More generally, tensor BDIs can be written as scalar
BDIs multiplied by symmetrized products of metric tensors plus surface terms.

Surface terms measure the failure of invariance under shifts of the loop
momentum. Indeed, changing the routing
\begin{equation}
    k\longrightarrow k+a
\end{equation}
in a divergent integral produces precisely integrals of total derivatives.
Consequently, momentum-routing invariance requires
\begin{equation}
    \Upsilon^{\mu_1\cdots\mu_r}=0.
    \label{eq:vanishing-surface-terms}
\end{equation}
In constrained IReg, surface terms are set to zero from the outset for
anomaly-free amplitudes. This ensures invariance under shifts of the loop
momenta and, in abelian gauge theories, is directly related to the
preservation of the Ward identities
\cite{FerreiraEtAl2012}. In anomalous amplitudes, by contrast, surface terms
may be retained until the physical Ward identities are imposed, allowing the
anomaly to be assigned consistently.

It is important to distinguish this procedure from naive symmetric
integration. IReg does not assume the equality between the terms in brackets in eq.
\eqref{eq:no-symmetric-integration}. Rather, it first identifies the
difference between the tensor and scalar integrals as a surface term and only
then fixes that term through momentum-routing invariance or the anomaly
conditions.

\subsection{Separation of ultraviolet and infrared scales}

The fictitious mass $\mu$ prevents the algebraic decomposition from generating
spurious infrared singularities. It is not the renormalization scale. The
genuine ultraviolet divergence is isolated by introducing a non-vanishing
scale $\lambda$ through the exact scale relation
\begin{equation}
    I_{\log}(\mu^2)
    =
    I_{\log}(\lambda^2)
    +
    b\ln\left(\frac{\lambda^2}{\mu^2}\right),
    \qquad
    b=\frac{i}{(4\pi)^2}.
    \label{eq:ireg-scale-relation}
\end{equation}
The object $I_{\log}(\lambda^2)$ contains the ultraviolet divergence and is
absorbed into the renormalization constants. The logarithm involving $\mu$
is retained together with the finite part of the amplitude.

For an amplitude that is infrared finite from the outset, the dependence on
$\mu$ cancels within the complete amplitude and the limit $\mu\to0$ can be
taken after the ultraviolet subtraction. In the present calculation, the
virtual and real contributions are separately infrared divergent. The same
mass $\mu$ is therefore retained in both sectors, and the limit $\mu\to0$ is
taken only after the renormalized virtual contribution has been combined with
the corresponding real-emission rate.

A minimal mass-independent subtraction prescription in IReg amounts to
subtracting only the BDIs evaluated at the renormalization scale, such as
$I_{\log}(\lambda^2)$. Renormalization-group functions can then be obtained
directly from their scale derivatives. At one loop,
\begin{equation}
    \lambda^2
    \frac{\partial I_{\log}(\lambda^2)}
         {\partial\lambda^2}
    =
    -b.
    \label{eq:bdi-scale-derivative}
\end{equation}

\subsection{Extension to multiloop amplitudes}

At multiloop order, the same procedure is applied recursively. Consider an
$N$-loop graph with loop momenta $k_1,\ldots,k_N$. When the integration in a
particular momentum $k_\ell$ is analyzed, all external momenta and all loop
momenta not yet integrated are collectively denoted by $q_i$. The propagators
then have the generic form
\begin{equation}
    \frac{1}{(k_\ell-q_i)^2-\mu^2}.
\end{equation}
Equation~\eqref{eq:ireg-propagator-identity} is applied with
$k\to k_\ell$ and $p_i\to q_i$. The value of $N_i$ must be chosen so that
the remainder is ultraviolet finite not only when $k_\ell\to\infty$ alone,
but in every ultraviolet region relevant to the graph.

After some of the loop integrations have been performed, logarithms generated
by lower-loop subgraphs appear in the remaining integrals. This motivates the
definition of higher-loop BDIs,
\begin{equation}
    I_{\log}^{(\ell)}(\mu^2)
    \equiv
    \int_{k_\ell}
    \frac{1}{(k_\ell^2-\mu^2)^2}
    \ln^{\ell-1}
    \left(
        -\frac{k_\ell^2-\mu^2}{\lambda^2}
    \right),
    \label{eq:higher-loop-bdi}
\end{equation}
and their tensor counterparts,
\begin{equation}
    I_{\log}^{(\ell)\,\mu_1\cdots\mu_{2r}}(\mu^2)
    \equiv
    \int_{k_\ell}
    \frac{
    k_\ell^{\mu_1}\cdots k_\ell^{\mu_{2r}}
    }
    {(k_\ell^2-\mu^2)^{r+2}}
    \ln^{\ell-1}
    \left(
        -\frac{k_\ell^2-\mu^2}{\lambda^2}
    \right).
    \label{eq:higher-loop-tensor-bdi}
\end{equation}
As at one loop, tensor higher-loop BDIs are reduced to scalar ones by setting
the corresponding total derivatives to zero. Generalized scale relations
separate $I_{\log}^{(\ell)}(\mu^2)$ into ultraviolet BDIs evaluated at
$\lambda^2$, lower-order BDIs, and powers of
$\ln(\mu^2/\lambda^2)$. Their scale derivatives satisfy, for $\ell\geq2$,
\begin{equation}
    \lambda^2
    \frac{\partial I_{\log}^{(\ell)}(\lambda^2)}
         {\partial\lambda^2}
    =
    -(\ell-1)I_{\log}^{(\ell-1)}(\lambda^2)
    -
    b(\ell-1)! .
    \label{eq:higher-bdi-scale-derivative}
\end{equation}

\subsubsection{Identification of subdivergences and the BPHZ structure}

The main difficulty at multiloop order is not merely to identify the overall
degree of divergence, but to distinguish all ultraviolet regions in which
one loop momentum, or a subset of loop momenta, becomes large. The recursive
IReg procedure can be summarized as follows:

\begin{enumerate}
    \item Identify the propagators that depend on physical external momenta
    or on loop momenta that are external to the subintegration under
    consideration.

    \item Apply the propagator identity
    \eqref{eq:ireg-propagator-identity} to these denominators, choosing the
    smallest values of $N_i$ that render the remainders ultraviolet finite
    in all relevant ultraviolet limits.

    \item Classify the resulting terms according to the different regions in
    which the internal momenta can approach infinity. This classification
    identifies the ultraviolet-divergent subgraphs of the original diagram.

    \item Bring each divergent term to normal form, reduce its tensor BDIs to
    scalar BDIs, and use the scale relations to express the ultraviolet
    content in terms of $I_{\log}^{(\ell)}(\lambda^2)$.

    \item Set aside the terms whose divergent factor has already been
    identified and repeat the procedure on the remaining loop integrations
    until all nested and overlapping ultraviolet regions have been
    exhausted.
\end{enumerate}

At the end of this procedure, two types of divergent structures are obtained:
\begin{equation}
    I_{\log}^{(\ell)}(\lambda^2)\,
    \mathcal J_{\mathrm{lower\;loop}},
    \qquad\text{and}\qquad
    I_{\log}^{(\ell)}(\lambda^2)\,
    P(p_1,\ldots,p_L),
    \label{eq:two-types-divergences}
\end{equation}
where $\mathcal J_{\mathrm{lower\;loop}}$ is a remaining loop integral and
$P$ is a polynomial in the external momenta and masses.

The first class contains the subdivergences. These are precisely the terms
removed by inserting the counterterms of the corresponding lower-loop
subgraphs, as prescribed by Bogoliubov's recursion formula. Once all
subdivergences have been subtracted, the second class contains the overall
divergence of the graph. Since it is a BDI multiplying a polynomial in
external momenta and masses, it is local and can be absorbed into the
counterterms allowed by the original Lagrangian.

The recursive IReg construction therefore does more than isolate divergent
integrals: it displays algebraically the counterterms selected by the BPHZ
forest structure. In this way, the method preserves the identity of a
renormalized subgraph when it is embedded into a higher-loop graph and
implements locality, Lorentz invariance, and perturbative unitarity without
explicitly evaluating the ultraviolet-divergent BDIs
\cite{CherchigliaSampaioNemes2011,AriasPerdomoEtAl2021}.

For the fermionic NNLO contribution studied here, this structure is directly
relevant. The quark vacuum-polarization insertion is first identified as a
ultraviolet-divergent subgraph and renormalized while preserving its
transverse tensor structure. Terms in which its BDI multiplies the remaining
vertex integral represent the associated subdivergence and are removed by
the lower-order counterterm. The divergence that remains after this
subtraction is the local overall divergence of the two-loop graph. The
auxiliary mass dependence that survives ultraviolet renormalization is
infrared in origin and cancels only after the virtual contribution is
combined with the corresponding real quark-pair channel.

%%%%%%%%%%%%%%

\section{The process \texorpdfstring{$Z\rightarrow q\bar q$}{Z to q qbar} in IReg}

\subsection{Virtual-virtual contributions}

We begin with the virtual-virtual contribution represented by diagram (a) from Fig.~\ref{fig:diagrams}. We denote by $k$ the outer-loop momentum and by $l$ the momentum circulating in the fermion bubble. The outgoing quark and antiquark momenta are $q$ and $\bar q$, respectively. Before collecting common factors, the two-loop vertex amplitude may be written as
\begin{align}
V_Z^{(2)}={}&\epsilon_\mu(z)\int_k
\bar u(q)(-ig_s\gamma^\alpha t^a)
\frac{-i(\slashed q+\slashed k)}{(q+k)^2}
\frac{-ie\gamma^\mu Z_-}{\sin(2\omega)}
\frac{i(\slashed{\bar q}-\slashed k)}{(\bar q-k)^2}
(-ig_s\gamma^\beta t^b)
\nonumber\\
&\times \frac{-iT_{\alpha\delta}\delta_{ac}}{k^2}
\big[i\Pi_f^{\delta\eta(cd)}(k)\big]
\frac{-iT_{\eta\beta}\delta_{db}}{k^2}
 v(\bar q),
\label{eq:explicit-two-loop-amplitude}
\end{align}
where the fermionic gluon polarization is
\begin{align}
i\Pi_f^{\delta\eta(cd)}(k)
={}&-\int_l\Tr\!\left[
(-ig_s\gamma^\delta t^c)\frac{i\slashed l}{l^2}
(-ig_s\gamma^\eta t^d)\frac{i(\slashed l-\slashed k)}{(l-k)^2}
\right].
\label{eq:explicit-fermion-polarization}
\end{align}
Throughout this section, Greek letters denote
Lorentz indices, whereas latin letters are color indices. As usual, the matrices $t^a$ are the generators of $SU(N_c)$ in the fundamental representation. The tensor
$T_{\rho\sigma}(k)$ is the transverse Lorentz projector associated with each
Landau-gauge gluon propagator and is defined explicitly below.

After performing the color trace and the Dirac trace of the fermion bubble, it is convenient to collect the common coupling and color factor as
\begin{equation}
\mathcal C_{\rm VV}=4\,\frac{i e\,N_fT_FC_F g_s^4}{\sin(2\omega)}.
\label{eq:global-factor}
\end{equation}
The amplitude then takes the compact form
\begin{align}
 V_Z^{(2)}
 &=\mathcal C_{\rm VV}\,\epsilon_\mu(z) \int_k\int_l
 \bar u(q)
 \frac{L_{\rm out}^{\alpha\beta\mu}L_{\rm in}^{\delta\eta}}{D_t}
 T_{\alpha\delta}T_{\beta\eta}
 v(\bar q),
 \label{eq:v2l}
\end{align}
with
\begin{align}
D_t&=k^4(q+k)^2(\bar q-k)^2l^2(l-k)^2,\\
L_{\rm out}^{\alpha\beta\mu}
&=\gamma^\alpha(\slashed q+\slashed k)\gamma^\mu Z_-(\slashed{\bar q}-\slashed k)\gamma^\beta,\\
L_{\rm in}^{\delta\eta}
&=2l^\delta l^\eta-g^{\delta\eta}(l^2-l\cdot k)-(k^\delta l^\eta+k^\eta l^\delta),\\
T_{\alpha\delta}&=g_{\alpha\delta}-\frac{k_\alpha k_\delta}{k^2}.
\end{align}
Here $N_f$ is the number of active quark flavors and $N_c$ is the number of
colors. The fundamental generators are normalized according to
$\Tr(t^a t^b)=T_F\delta^{ab}$, with $T_F=1/2$. The quantity
$C_F=(N_c^2-1)/(2N_c)$ is the quadratic Casimir invariant of the fundamental
representation. Finally, $g_s$ denotes
the strong gauge coupling. The numerical factor $4$ in Eq.~\eqref{eq:global-factor} originates from the Dirac trace of the fermion bubble. The  neutral-current couplings are written as
\begin{equation}
Z_\pm=g_V\pm g_A\gamma_5,
\qquad
g_V=I_3-2Q_q\sin^2\omega,
\qquad
g_A=I_3,
\label{eq:Zpm-def}
\end{equation}
where $I_3$ is the third component of weak isospin, $Q_q$ is the quark electric charge in units of $e$, and $\omega$ is the weak mixing angle. Landau gauge is used for the internal gluon propagator.

We will focus first on the inner-loop object
\begin{equation}
 \int_l\frac{L_{\rm in}^{\delta\eta}}{l^2(l-k)^2}
\end{equation}
which is basically the quark-loop contribution to the gluon polarization tensor at one-loop. We notice that the last structure in $L_{\rm in}^{\delta\eta}$ vanishes after contraction with the transverse projectors,
\begin{equation}
 (k^\delta l^\eta+k^\eta l^\delta)T_{\alpha\delta}T_{\beta\eta}=0.
 \label{eq:mix-vanish}
\end{equation}
Writing the reduced term $\bar L_{\rm in}^{\delta\eta}=2l^\delta l^\eta-g^{\delta\eta}(l^2-l\cdot k)$, one obtains
\begin{align}
\label{eq:Lbar}
 \bar L_{\rm in}^{\delta\eta}T_{\alpha\delta}T_{\beta\eta}
 &=\bar L_{\rm in}^{\alpha\beta}+\Lambda^{\alpha\beta},\quad\mbox{where}\\
\Lambda^{\alpha\beta}
&=\frac{k^\alpha k^\beta}{k^2}(l^2-l\cdot k)
 -\frac{2l\cdot k}{k^2}(k^\alpha l^\beta+l^\alpha k^\beta)
 +\frac{2k^\alpha k^\beta}{k^4}(k\cdot l)^2 .
 \label{eq:Lambda}
\end{align}
As we are going to see, the term $\Lambda^{\alpha\beta}$ is essential. If only the metric contractions are kept ($\bar L_{\rm in}^{\alpha\beta}$), a non-transverse contribution is generated at intermediate stages. The projector pieces remove it and yield the gauge-invariant polarization insertion. 

Regarding the integrals, we will employ IReg to disentangle the UV divergent part. After this step is achieved, the remaining integrals will be rewritten using Feynman parametrization. We will not proceed further for each integral, only performing the integral over the intermediate parameter introduced in the Feynman parametrization in the end. Defining each UV-divergent integral associated with the inner-loop momentum $l$ as $\mathit J=\operatorname{fin}(\mathit J)+\operatorname{div}(\mathit J)$, with the finite and divergent pieces separated according to the IReg prescription, one obtains for the finite parts
\begin{align}
\mbox{fin}\int_l\frac{l_\alpha l_\beta}{l^2(l-k)^2}
 &=\frac{i}{(4\pi)^2}\int_0^1\!\dd x\,
   \frac{k^2 k_\alpha k_\beta C_1(x)+g_{\alpha\beta}k^4 C_2(x)}{D_0},
\label{eq:tens}\\
\mbox{fin}\int_l\frac{l^2}{l^2(l-k)^2}
 &=\frac{i k^4}{(4\pi)^2}\int_0^1\!\dd x\,\frac{C_0(x)}{D_0},\\
\mbox{fin}\int_l\frac{(l\cdot k)^2}{l^2(l-k)^2}
 &=\frac{i k^6}{(4\pi)^2}\int_0^1\!\dd x\,\frac{C^{'}_0(x)}{2D_0},\\
\mbox{fin}\int_l\frac{l\cdot k}{l^2(l-k)^2}
 &=\frac{i k^4}{(4\pi)^2}\int_0^1\!\dd x\,\frac{C^{'}_0(x)}{D_0},
\end{align}
where the denominator is defined as 
\begin{equation}
D_0=k^2-\frac{\mu^2}{x(1-x)}.
\end{equation}
A mass $\mu^{2}$ was introduced in each propagator, as usual in the definition of IReg, in order to parametrize the IR divergent behavior. The $C_{i}(x)$ are given by
\begin{align}\label{coefX}
C_1(x)&=\frac{(-1+2x)(1+x+x^2)}{3x},\quad
C_2(x)=\frac{1-7x(1-2x)-8x^3}{12x},\\
C_0(x)&=-\frac{1}{2}(4x^2-1),\quad
C'_0(x)=-\frac{1}{2}\left(\frac{1}{x}-2\right).
\end{align}
Using the $x\leftrightarrow 1-x$ symmetry under an integral whose remaining
weight is symmetric, the combinations which appear in the amplitude can be
represented by
\begin{equation}
F(x)\equiv-\frac{4}{3}\big(C_0(x)+C'_0(x)\big)
 =\frac{2}{3}\left(\frac{1}{x}-3+4x^2\right).
 \label{eq:Fdef}
\end{equation}
Similarly, we have the following relations that are equivalent under integration
\begin{equation}
 -2(C'_0+2C_2)\overset{x\leftrightarrow1-x}{\sim}
 -4(C_1+2C_2)\overset{x\leftrightarrow1-x}{\sim}
 4(2C_2-C_0)\overset{x\leftrightarrow1-x}{\sim}
 -\frac{4}{3}(C_0+C'_0)=F(x).\label{eq:coeFx}
\end{equation}
On the other hand, the corresponding divergent parts are given by
\begin{equation}
\label{tensdiv}
    \begin{array}{cc}
     \mbox{div}   \int_l \frac{l_\alpha l_\beta}{l^2(l-k)^2} = -\frac{1}{3} k^2 (\frac{g_{\alpha\beta}}{4} -\frac{{k_\alpha} k_{\beta}}{k^2}) I_{\log}(\mu^{2})
    \end{array}
\end{equation} 
\begin{equation}
\label{lsk2div}
    \mbox{div} \int_l \frac{(l \cdot k)^2}{l^2(l-k)^2} = \frac{k^4}{4} I_{\log}(\mu^{2})
\end{equation}
\begin{equation}
\label{Bdiv}
    \mbox{div} \int_l \frac{1}{l^2(l-k)^2} = I_{\log}(\mu^{2})
\end{equation}
\begin{equation}
\mbox{div} \int_l \frac{l^2}{l^2(l-k)^2}=0
\end{equation}
\begin{equation}
   \mbox{div}\int_l \frac{l \cdot k}{l^2(l-k)^2} = \frac{k^2}{2 } I_{\log}(\mu^{2})   
\end{equation}
where, in the quadratic divergent integrals the quadratic BDI is set to zero,  $I_{\mathrm{quad}}(\mu^{2})=0$.

It is fruitful to establish a ``substitution rule" to convert the terms containing the $C_i$ coefficients of the finite integrals to the corresponding divergent ones

\begin{align}
\frac{i}{(4\pi)^2}\int_0^1\!\dd x\,\frac{C_1}{D_0}
&\longrightarrow C_1^{\rm UV}=\frac{1}{3k^2}I_{\log}(\mu^2),\\
\frac{i}{(4\pi)^2}\int_0^1\!\dd x\,\frac{C_2}{D_0}
&\longrightarrow C_2^{\rm UV}=-\frac{1}{12k^2}I_{\log}(\mu^2),\\
\frac{i}{(4\pi)^2}\int_0^1\!\dd x\,\frac{C'_0}{D_0}
&\longrightarrow {C'_0}^{\rm UV}=\frac{1}{2k^2}I_{\log}(\mu^2),\\
\frac{i}{(4\pi)^2}\int_0^1\!\dd x\,\frac{C_0}{D_0}
&\longrightarrow C_0^{\rm UV}=0,\\
\frac{i}{(4\pi)^2}\int_0^1\!\dd x\,\frac{F(x)}{D_0}
&\longrightarrow F^{\rm UV}=-\frac{2}{3k^2}I_{\log}(\mu^2).
\label{fd}
\end{align}
These substitutions satisfy relations analogous to Eq.~\eqref{eq:coeFx},

\begin{equation}
\label{coeFxdiv}
    \begin{array}{cc}
        -2({C'_0}^{UV}+2C^{UV}_2) = -4 (C_1^{UV}+2C_2^{UV}) = 4 (2C^{UV}_2-C^{UV}_0) = -\frac{4}{3} ({C_0}^{UV}+{C'_0}^{UV}) \\
        -2(C^{UV}_1+C^{UV}_0) = -\frac{2}{3k^2} I_{\log}(\mu^{2}) \equiv F^{UV} \
    \end{array}
\end{equation}

Since these latter coefficients are expressed solely in terms of $I_{\log}(\mu^{2})$ in Lorentz structures identical to the finite contributions, it is clear that the amplitude obtained from the $\mbox{div} J$ contributions are identical to the counterterms necessary for the inner one-loop renormalization. In other words the amplitude obtained from the UV finite inner loop contributions $\mbox{fin} J$ provides all what is necessary to analyze and obtain the two-loop order amplitude and renormalization counterterms. The latter emerge as divergent outer loop integrals stemming from $\mbox{fin} J$ contributions.

The polarization tensor, with the color and flavor factors already extracted into Eq.~\eqref{eq:global-factor}, is transverse as  expected \footnote{Taking only the metric pieces of the projectors into account in Eq.~\ref{eq:Lbar} generates a non-transverse contribution $k^2 k^\alpha k^\beta \int_{0}^{1} dx \frac{2 C'_0}{D_0}+ k^\alpha k^\beta I_{log}$, which is cancelled by the contribution from $\Lambda^{\alpha\beta}$.} 
\begin{equation}
 i\Pi^{\alpha\beta}(k)= -4\int_l\frac{L_{\rm in}^{\delta\eta}}{l^2(l-k)^2}T_{\alpha\delta}T_{\beta\eta}
= -4\left\{\frac{k^2}{3}I_{\log}(\mu^2)-\frac{i k^4}{(4\pi)^2}\int_0^1\!\dd x\,\frac{F(x)}{2D_0}\right\}
\left(g^{\alpha\beta}-\frac{k^\alpha k^\beta}{k^2}\right).
\label{eq:PiGI}
\end{equation}
After the $x$ integration one recovers the standard IReg form
\begin{equation}
 i\Pi^{\alpha\beta}(k)=
 -\frac{4k^2}{3}\left[I_{\log}(\mu^2)+b\left(\frac{5}{3}+\ln\left(-\frac{\mu^2}{k^2}\right)\right)\right]
 \left(g^{\alpha\beta}-\frac{k^\alpha k^\beta}{k^2}\right),
 \qquad b=\frac{i}{(4\pi)^2}.
\label{eq:PiGI-integrated}
\end{equation}
The BDI part of Eq.~\eqref{eq:PiGI} is precisely the one-loop polarization subgraph counterterm.  In the BPHZ scheme it is subtracted before the remaining finite insertion is integrated over the outer loop.

We now insert the finite part of Eq.~\eqref{eq:PiGI} into the outer loop. For compactness, the factor of $i$ carried by the finite inner-loop contribution is displayed together with $F(x)$ in the following formulas.
\begin{align}
 V_Z^{(2)}
 &=- \epsilon_\mu \frac{\mathcal C_{\rm VV}}{(4\pi)^2}\int_k
 \frac{\bar u(q)\gamma^\alpha(\slashed q+\slashed k)\gamma^\mu Z_-
 (\slashed{\bar q}-\slashed k)\gamma^\beta v(\bar q)}
 {(q+k)^2(\bar q-k)^2}
 \int_0^1\!\dd x\,
 \frac{iF(x)}{2D_0}
 \left(g_{\alpha\beta}-\frac{k_\alpha k_\beta}{k^2}\right).
\end{align}

For simplicity, we will define $\Delta=q-\bar q$. We use a fixed right- or left-most ordering prescription  \cite{tsai2009advantage,tsai2011gauge} for $\gamma_5$ in the open fermionic string before applying the massless equations of motion. After performing the Dirac algebra, we notice that terms proportional to $\Delta^\mu\slashed{k}$ will be proportional to $\slashed{q}-\slashed{\bar{q}}$, thus vanishing after the usage of on-shell relations. Therefore, we obtain
\begin{align}
V_Z^{(2)}
={}& {\epsilon_\mu \bar u}(q)
 \frac{\mathcal C_{\rm VV}}{(4\pi)^2}\int_{k,x}\left(
 \frac{\gamma^\mu Z_-}{2k^2}-\frac{2k^\mu\slashed{k}Z_-}{k^4}
 \right)\frac{iF(x)}{D_0}\,v(\bar q)
\\
&+\epsilon_\mu {\bar u}(q)\frac{\mathcal C_{\rm VV}}{(4\pi)^2} 
 \int_{k,x}\frac{\left((-2q\cdot\bar q+k\cdot\Delta\right)k^{2}\gamma^\mu +2k\cdot\Delta k^\mu\slashed{k})Z_-}
 {D_0k^2(q+k)^2(\bar q-k)^2}
 iF(x)\,v(\bar q).
\end{align}
The first line contains only BDI's. Actually, the combination is proportional to a surface term, rendering a null result. Thus, we are left only with the second line.

It is useful to introduce the set of integrals 
\begin{align}
A_q^\mu&=-\int_{k,x}\frac{k^\mu\slashed{k}\,k\cdot q}{D_0k^2(q+k)^2(\bar q-k)^2}iF(x),
&
A_{\bar q}^\mu&=\int_{k,x}\frac{k^\mu\slashed{k}\,k\cdot\bar q}{D_0k^2(q+k)^2(\bar q-k)^2}iF(x),\\
H_q&=-\int_{k,x}\frac{k\cdot q}{D_0(q+k)^2(\bar q-k)^2}iF(x),
&
H_{\bar q}&=\int_{k,x}\frac{k\cdot\bar q}{D_0(q+k)^2(\bar q-k)^2}iF(x),\\
G_q&=\int_{k,x}\frac{(k\cdot q)^2}{D_0(q+k)^2(\bar q-k)^2}iF(x),
&
I&=\int_{k,x}\frac{1}{D_0(q+k)^2(\bar q-k)^2}i F(x).
\end{align}
In the massless limit,
\begin{equation}
 A_q^\mu=A_{\bar q}^\mu=
 \left(\frac{H_q}{2}+\frac{G_q}{2q\cdot\bar q}\right)\gamma^\mu,
\label{eq:Aq}
\end{equation}
where the divergent part of $G_q$ vanishes because it is proportional to $q^2$.  The finite virtual contribution can then be written as
\begin{equation}
\left. V_Z^{(2)}\right|_{\rm fin}
=\frac{16 i e\,N_f C_FT_F\alpha_s^2}{\sin(2\omega)} \epsilon_\mu\bar u(q)\gamma^\mu Z_-v(\bar q)
\left[\frac{G_q}{2q\cdot\bar q}-H_q+\frac{q\cdot\bar q}{2}I\right]
.
\label{eq:Vfin}
\end{equation}
The two-loop correction has the same external neutral-current Dirac structure as the Born amplitude,
\begin{equation}
\mathcal M_Z^{(0)}=-\frac{i e}{\sin(2\omega)}\,
\epsilon_\mu(z)\,\bar u(q)\gamma^\mu Z_-v(\bar q).
\label{ta0}
\end{equation}
Thus the two-loop result may be viewed as the Born current multiplied by a scalar form-factor correction.  The scalar form factor is identical to the one resulting from a pure vector-current in the comparison to Ref.~\cite{Page:2018ljf}.

\subsection{UV renormalization}

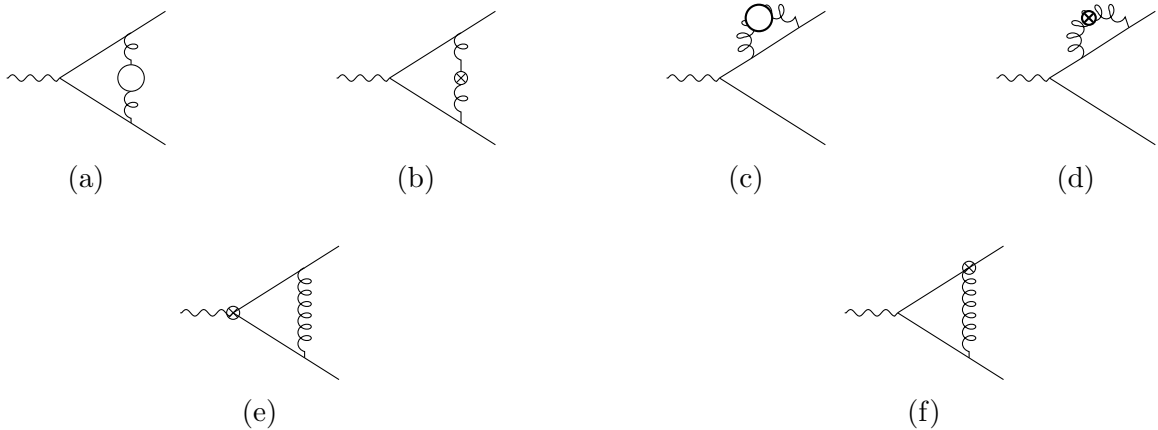
\begin{figure}[htbp]
\centering

% ---------- (a) ----------
\begin{subfigure}{0.23\textwidth}
\centering
\begin{tikzpicture}[scale=0.7, transform shape]
\begin{feynman}
  \vertex (a)  at (-1.0,0);
  \vertex (v)  at (0,0);

  \vertex (u1) at (1.35,0.85);
  \vertex (u2) at (2.,1.26);

  \vertex (d1) at (1.35,-0.85);
  \vertex (d2) at (2.,-1.26);

  \vertex (gup) at (1.35,0.25);
  \vertex (gdn) at (1.35,-0.25);
  \vertex (c)   at (1.35,0);

  \diagram*{
    (a) -- [photon] (v),
    (v) -- [plain] (u1) -- [plain] (u2),
    (d2) -- [plain] (d1) -- [plain] (v),
    (u1) -- [gluon] (gup),
    (gdn) -- [gluon] (d1),
  };

  \draw[fill=white] (c) circle [radius=0.25];
\end{feynman}
\end{tikzpicture}
\caption{}
\label{fig.a}
\end{subfigure}
\hfill
% ---------- (b) ----------
\begin{subfigure}{0.23\textwidth}
\centering
\begin{tikzpicture}[scale=0.7, transform shape]
\begin{feynman}
  \vertex (a)  at (-1.0,0);
  \vertex (v)  at (0,0);

  \vertex (u1) at (1.35,0.85);
  \vertex (u2) at (2.,1.26);

  \vertex (d1) at (1.35,-0.85);
  \vertex (d2) at (2.,-1.26);

  \vertex [crossed dot] (c) at (1.35,0) {}; 

  \diagram*{
    (a) -- [photon] (v),
    (v) -- [plain] (u1) -- [plain] (u2),
    (d2) -- [plain] (d1) -- [plain] (v),
    (u1) -- [gluon] (c),
    (c) -- [gluon] (d1),
  };
\end{feynman}
\end{tikzpicture}
\caption{}
\label{fig.b}
\end{subfigure}
\hfill
% ---------- (c) ----------
\begin{subfigure}{0.23\textwidth}
\centering
\begin{tikzpicture}[scale=0.7, transform shape]
\begin{feynman}
  \vertex (a)  at (-1.0,0);
  \vertex (v)  at (0,0);

  \vertex (u1) at (.6,0.37);
  \vertex (u2) at (1.5,0.95);
  \vertex (u3) at (2.,1.26);

  \vertex (d2) at (2.,-1.26);

  \diagram*{
    (a) -- [photon] (v),
    (v) -- [plain] (u1) -- [plain] (u2) -- [plain] (u3),
    (d2) -- [plain] (v),
  };

  \draw[gluon, half left, looseness=1.8]
    (u1) to coordinate[pos=0.5] (c) (u2);

  \draw[fill=white, line width=0.8pt] (c) circle [radius=0.25];
\end{feynman}
\end{tikzpicture}
\caption{}
\label{fig.c}
\end{subfigure}
\hfill
% ---------- (d) ----------
\begin{subfigure}{0.23\textwidth}
\centering
\begin{tikzpicture}[scale=0.7, transform shape]
\begin{feynman}
  \vertex (a)  at (-1.0,0);
  \vertex (v)  at (0,0);

  \vertex (u1) at (.6,0.37);
  \vertex (u2) at (1.5,0.95);
  \vertex (u3) at (2.,1.26);

  \vertex (d2) at (2.,-1.26);

  \diagram*{
    (a) -- [photon] (v),
    (v) -- [plain] (u1) -- [plain] (u2) -- [plain] (u3),
    (d2) -- [plain] (v),
  };

  \draw[gluon, half left, looseness=1.8]
    (u1) to coordinate[pos=0.5] (c) (u2);

  \draw[fill=white, line width=0.8pt] (c) circle [radius=0.13];

  \draw[line width=0.8pt]
    ($(c)+(-0.08,-0.08)$) -- ($(c)+(0.08,0.08)$);
  \draw[line width=0.8pt]
    ($(c)+(-0.08,0.08)$) -- ($(c)+(0.08,-0.08)$);
\end{feynman}
\end{tikzpicture}
\caption{}
\label{fig.d}
\end{subfigure}

\vspace{0.6cm}

% ---------- (e) ----------
\begin{subfigure}{0.5\textwidth}
\centering
\begin{tikzpicture}[scale=0.7, transform shape]
\begin{feynman}
  \vertex (a)  at (-1.0,0);
  \vertex (v)  at (0,0);

  \vertex (u1) at (1.35,0.85);
  \vertex (u2) at (2.,1.26);

  \vertex (d1) at (1.35,-0.85);
  \vertex (d2) at (2.,-1.26);

  \vertex [crossed dot] (c) at (0,0) {}; 

  \diagram*{
    (a) -- [photon] (v),
    (v) -- [plain] (u1) -- [plain] (u2),
    (d2) -- [plain] (d1) -- [plain] (v),
    (u1) -- [gluon] (d1),
  };
\end{feynman}
\end{tikzpicture}
\caption{}
\label{fig.e}
\end{subfigure}
\hfill
% ---------- (f) ----------
\begin{subfigure}{0.45\textwidth}
\centering
\begin{tikzpicture}[scale=0.7, transform shape]
\begin{feynman}
  \vertex (a)  at (-1.0,0);
  \vertex (v)  at (0,0);

  \vertex (u1) at (1.35,0.85);
  \vertex (u2) at (2.,1.26);

  \vertex (d1) at (1.35,-0.85);
  \vertex (d2) at (2.,-1.26);

  \diagram*{
    (a) -- [photon] (v),
    (v) -- [plain] (u1) -- [plain] (u2),
    (d2) -- [plain] (d1) -- [plain] (v),
    (u1) -- [gluon] (d1),
  };

  \node[crossed dot] at (u1) {};
\end{feynman}
\end{tikzpicture}
\caption{}
\label{fig.f}
\end{subfigure}
\hfill

\caption{Feynman diagrams relevant for the renormalization of the NNLO $Z\rightarrow q\bar{q}$ process.}
\label{fig:diagrams}
\end{figure}

In this section we discuss the UV renormalization of $Z\to q\bar q$ at NNLO, restricting attention to the contribution proportional to $N_f$. We use the same on-shell organization as in the NLO analysis. At this order, the diagrams proportional to $N_f$ are those that contain an internal fermion bubble, namely Figs.~\ref{fig.a} and \ref{fig.c}; an analogous external-leg diagram with the bubble on the antiquark line is understood. The corresponding one-loop counterterm insertions are shown in Figs.~\ref{fig.b} and \ref{fig.d}.

The sum of the two-loop vertex graph in Fig.~\ref{fig.a} and the subgraph counterterm in Fig.~\ref{fig.b} is precisely the combination constructed in the previous subsection. The fermion-bubble subdivergence is removed locally, and no additional UV divergence remains in this $N_f$ contribution to the renormalized neutral-current vertex.

The external-leg contribution is treated in the same way. For the diagrams in Figs.~\ref{fig.c} and \ref{fig.d}, the fermion-bubble BDI is subtracted before the remaining outer-loop integration is performed. As shown explicitly in Appendix~\ref{app:wfr}, the UV-divergent part of the resulting wave-function contribution vanishes. Consequently, the first four diagrams in Fig.~\ref{fig:diagrams} generate no additional local UV counterterm in the $N_f$ contribution considered here.

The counterterm diagram in Fig.~\ref{fig.e} is proportional to the renormalization of the electroweak coupling. Although that renormalization constant may contain an $N_f$ contribution, this term belongs to a different coupling order, proportional to $\alpha_e\alpha_s$, rather than to the $\alpha_s^2$ correction isolated in the present calculation. It is therefore outside the contribution considered here.

Finally, Fig.~\ref{fig.f} is associated with the quark--gluon vertex renormalization constant. At the order displayed it does not contain the required $N_f$ term and therefore does not contribute to the color--flavor structure studied here. Taken together, these observations show that, after subtraction of the fermion-bubble subdivergence, no additional UV-divergent counterterm remains in the NNLO contribution proportional to $C_FT_FN_f$.

This conclusion can also be checked through the Ward-identity relation between the neutral-current vertex and the quark self-energy. In the $N_f$ contribution, the relevant two-loop self-energy is again the diagram containing a fermion bubble. Appendix~\ref{app:self} shows that its UV-divergent combination vanishes after the corresponding subgraph subtraction, consistently with the vertex analysis above.

\subsection{Real contributions}

The real contribution associated with the $N_f$ part of the NNLO correction is the four-particle final state
\begin{equation}
  V(p)\longrightarrow q(q)+\bar q(\bar q)+q_f(k)+\bar q_f(\bar k),
  \qquad p=q+\bar q+k+\bar k,
\end{equation}
where $V$ denotes the external neutral current. The flavor $f$ is summed over the $N_f$ quarks that may be produced by the splitting of the internal gluon. 

To extract the IR singularities from the real contribution, we follow the prescription used in Ref.~\cite{Pereira:2022mne}. The Dirac algebra is first performed with massless quarks, so that the squared matrix element can be reduced entirely to Lorentz invariants. The phase-space integration is then carried out with the fictitious mass assigned to the particles associated, through the Cutkosky cut, with the fermion bubble of the virtual contribution. In the present case,
\begin{equation}
q^2=\bar q^{\,2}=0,
\qquad
k^2=\bar k^{\,2}=\mu^2,
\qquad
s=p^2.
\label{eq:real-mass-assignment}
\end{equation}
For the cut fermion pair, the mass-dependent part of the spin completeness relation is retained until the phase-space integration has been performed. This is the real-emission counterpart of using the same fictitious mass in the virtual bubble and makes the Cutkosky correspondence between the two sectors explicit. Within this prescription, 
the IR divergences will be parametrized in terms of the fictitious mass $\mu^{2}$. As we are going to show, the regulator dependence precisely cancels after the real contribution is combined with the renormalized virtual contribution.

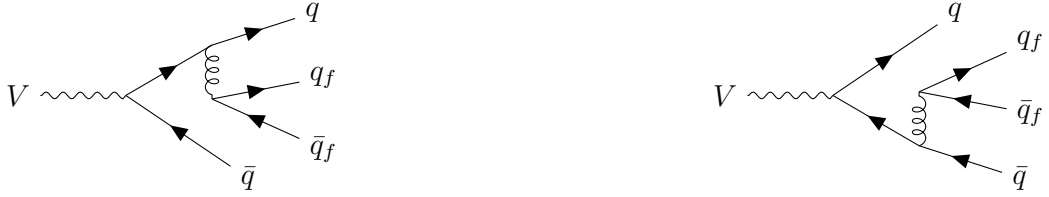
\begin{figure}[htbp]
\centering
\begin{subfigure}{0.45\textwidth}
\centering
\begin{tikzpicture}[scale=0.95, transform shape]
\begin{feynman}
  \vertex (v) at (-1.5,0) {$V$};
  \vertex (a) at (0,0);
  \vertex (b) at (1.2,0.7);
  \vertex (qout) at (2.6,1.15) {$q$};
  \vertex (qbout) at (1.7,-1.15) {$\bar q$};
  \vertex (g) at (1.2,-0.05);
  \vertex (kout) at (2.75,0.25) {$q_f$};
  \vertex (kbout) at (2.75,-0.75) {$\bar q_f$};
  \diagram*{
    (v) -- [photon] (a),
    (a) -- [fermion] (b) -- [fermion] (qout),
    (qbout) -- [fermion] (a),
    (b) -- [gluon] (g),
    (g) -- [fermion] (kout),
    (kbout) -- [fermion] (g),
  };
\end{feynman}
\end{tikzpicture}
\caption{Emission from the quark line.}
\label{fig:real-q}
\end{subfigure}
\hfill
\begin{subfigure}{0.45\textwidth}
\centering
\begin{tikzpicture}[scale=0.95, transform shape]
\begin{feynman}
  \vertex (v) at (-1.5,0) {$V$};
  \vertex (a) at (0,0);
  \vertex (qout) at (1.7,1.15) {$q$};
  \vertex (b) at (1.2,-0.7);
  \vertex (qbout) at (2.6,-1.15) {$\bar q$};
  \vertex (g) at (1.2,0.05);
  \vertex (kout) at (2.75,0.75) {$q_f$};
  \vertex (kbout) at (2.75,-0.25) {$\bar q_f$};
  \diagram*{
    (v) -- [photon] (a),
    (a) -- [fermion] (qout),
    (qbout) -- [fermion] (b) -- [fermion] (a),
    (b) -- [gluon] (g),
    (g) -- [fermion] (kout),
    (kbout) -- [fermion] (g),
  };
\end{feynman}
\end{tikzpicture}
\caption{Emission from the antiquark line.}
\label{fig:real-qbar}
\end{subfigure}
\caption{Real-real diagrams contributing to the $N_f$ part of the inclusive NNLO correction.  The quark pair $q_f\bar q_f$ is unresolved and is integrated over the four-particle phase space.}
\label{fig:real-diagrams}
\end{figure}

It is useful to introduce the momentum assignment
\begin{equation}
  k_1\equiv q,
  \qquad k_2\equiv \bar q,
  \qquad k_3\equiv k,
  \qquad k_4\equiv \bar k,
\end{equation}
with invariants
\begin{equation}
  s_{ij}=(k_i+ k_j)^2,
  \qquad
  s_{ijk}=(k_i+k_j+k_k)^2,
  \qquad
  s=(k_1+k_2+k_3+k_4)^2 .
\end{equation}
The denominators that appear in the real matrix element are
\begin{equation}
  s_{34}=(k+\bar k)^2,
  \qquad
  s_{134}=(q+k+\bar k)^2,
  \qquad
  s_{234}=(\bar q+k+\bar k)^2 .
\end{equation}
In the notation of the virtual part, the real channel is the cut counterpart of the fermion-bubble insertion in the gluon propagator.  The two tree-level amplitudes differ only by the emission of the off-shell gluon from the quark or from the antiquark line.  They are shown in Fig.~\ref{fig:real-diagrams}.

We write the neutral-current vertex as
\begin{equation}
  \GammaV^\mu=\gamma^\mu Z_- ,
  \qquad
  Z_-=g_V-g_A\gamma_5,
  \qquad
  Z_+=g_V+g_A\gamma_5 .
\end{equation}
For a purely electromagnetic vector current one sets $g_V=1$ and $g_A=0$.  The polarization sum is written as
\begin{equation}
  \sum_\lambda \epsilon^{(\lambda)\,*}_\mu(p)\epsilon^{(\lambda)}_\nu(p)
  =-T_{\mu\nu}(p),
  \qquad
  T_{\mu\nu}(p)=g_{\mu\nu}-\frac{p_\mu p_\nu}{p^2} .
\end{equation}
For massless external quarks the longitudinal part drops from the inclusive vector-current projection after the two real diagrams are combined.  Keeping $T_{\mu\nu}$ in intermediate steps is nevertheless useful as a check of current conservation.

With these conventions, the quark-emission amplitude is
\begin{align}
  \M_q
  &=(-\ii)^5\frac{eg_s^2}{\sin(2\omega)}
    \epsilon_\mu(p)
    \bar u(q) \gamma^\alpha t^a
    \frac{\slashed q+\slashed k+\slashed{\bar k}}{(q+k+\bar k)^2}
    \GammaV^\mu v(\bar q)
    \frac{g_{\alpha\beta}}{(k+\bar k)^2}
    \bar u(k)\gamma^\beta t^a v(\bar k) .
    \label{eq:Mq-real}
\end{align}
Using momentum conservation, $(q+k+\bar k)=p-\bar q$, this may be written in the compact form
\begin{equation}
  \M_q
  =(-\ii)^5\frac{eg_s^2}{\sin(2\omega)}
  \epsilon_\mu(p)
  \mathcal R_q^{\mu\delta}
  \frac{p_\delta-\bar q_\delta}{s_{134} s_{34}}%\,k\cdot\bar k},
  \label{eq:Mq-compact}
\end{equation}
where
\begin{equation}
  \mathcal R_q^{\mu\delta}
  =\big[\bar u(q)Z_+t^a\gamma^\alpha\gamma^\delta\gamma^\mu v(\bar q)\big]
   \big[\bar u(k)\gamma_\alpha t^a v(\bar k)\big] .
\end{equation}
The antiquark-emission amplitude is obtained by exchanging $q\leftrightarrow\bar q$ and reversing the corresponding fermion-line ordering,
\begin{equation}
  \M_{\bar q}
  =(-\ii)^5\frac{eg_s^2}{\sin(2\omega)}
  \epsilon_\mu(p)
  \mathcal R_{\bar q}^{\mu\delta}
  \frac{p_\delta-q_\delta}{s_{234} s_{34}}%\,k\cdot\bar k},
  \label{eq:Mqb-compact}
\end{equation}
with
\begin{equation}
  \mathcal R_{\bar q}^{\mu\delta}
  =-\big[\bar u(q)Z_+t^a\gamma^\mu\gamma^\delta\gamma^\alpha v(\bar q)\big]
    \big[\bar u(k)\gamma_\alpha t^a v(\bar k)\big] .
\end{equation}
The minus sign is the relative sign associated with emission from the antiquark line.  The squared real matrix element is then organized as
\begin{equation}
  |\M_q+\M_{\bar q}|^2
  =|\M_q|^2+|\M_{\bar q}|^2
  +2\Ree(\M_q^*\M_{\bar q}) .
  \label{eq:real-squared-decomp}
\end{equation}

The spin sums reduce the building blocks in Eq.~\eqref{eq:real-squared-decomp} to products of two traces.  For the quark-emission term one obtains
\begin{align}
  \mathcal R_q^{\dagger\nu\eta}\mathcal R_q^{\mu\delta}
  &=\Tr\!\left[
      \gamma^\nu\gamma^\eta\gamma^\beta t^bZ_-\slashed q
      Z_+t^a\gamma^\alpha\gamma^\delta\gamma^\mu\slashed{\bar q}
    \right]
    \Tr\!\left[t_b\gamma_\beta\slashed k\gamma_\alpha t_a\slashed{\bar k}\right] .
    \label{eq:AqAq-trace}
\end{align}
The antiquark-emission term follows by $q\leftrightarrow\bar q$ and $s_{134}\leftrightarrow s_{234}$.  The interference term is
\begin{align}
  \mathcal R_q^{\dagger\nu\eta}\mathcal R_{\bar q}^{\mu\delta}
  &=-\Tr\!\left[
      \gamma^\nu\gamma^\eta\gamma^\beta t^bZ_-\slashed q
      Z_+t^a\gamma^\mu\gamma^\delta\gamma^\alpha\slashed{\bar q}
    \right]
    \Tr\!\left[t_b\gamma_\beta \slashed k \gamma_\alpha t_a \slashed{\bar k}\right] .
    \label{eq:AqAqb-trace}
\end{align}
The unresolved pair trace is
\begin{equation}
  L_{\alpha\beta}=\Tr(\slashed k\gamma_\alpha\slashed{\bar k}\gamma_\beta)
  =4\left(k_\alpha\bar k_\beta+k_\beta\bar k_\alpha-g_{\alpha\beta} k\cdot\bar k\right).
  \label{eq:unresolved-trace}
\end{equation}
The terms proportional to $g_Vg_A$ contain one $\gamma_5$.  In the inclusive vector projection considered here, they vanish after the massless equations of motion and the symmetric phase-space integration are used.  The real correction is therefore proportional to $g_V^2+g_A^2$, and in the electromagnetic case this factor is equal to one.

With these conventions the squared contributions can be written as
\begin{align}
  |\M_q|^2
  &=-\mathcal{N}_{\rm real}
  T_{\mu\nu}(p)
  \mathcal R_q^{\dagger\nu\eta}\mathcal R_q^{\mu\delta}
  \frac{(p-\bar q)_\delta(p-\bar q)_\eta}{s_{134}^2s_{34}^2},
  \label{eq:Mq2-general}\\
  |\M_{\bar q}|^2
  &=-\mathcal{N}_{\rm real}
  T_{\mu\nu}(p)
  \mathcal R_{\bar q}^{\dagger\nu\eta}\mathcal R_{\bar q}^{\mu\delta}
  \frac{(p-q)_\delta(p-q)_\eta}{s_{234}^2s_{34}^2},
  \label{eq:Mqb2-general}\\
  2\Ree(\M_q^*\M_{\bar q})
  &=-2\mathcal{N}_{\rm real}
  \Ree\left[
  T_{\mu\nu}(p)
  \mathcal R_q^{\dagger\nu\eta}\mathcal R_{\bar q}^{\mu\delta}
  \frac{(p-q)_\delta(p-\bar q)_\eta}{s_{134}s_{234} s_{34}^2}
\right].
  \label{eq:interference-general}
\end{align}
For simplicity we defined the common overall factor
\begin{equation}
  \mathcal{N}_{\rm real}
  =\frac{e^2g_s^4}{\sin^2(2\omega)}\,C_FT_FN_fN_c.
\end{equation}
Within this definition, the color matrices inside $R_{x}$ are understood to be omitted. 
Equations~\eqref{eq:Mq2-general}--\eqref{eq:interference-general} are the most useful starting point for checking the real calculation algebraically, because they keep the correspondence with the cut fermion-bubble contribution manifest.

To proceed further, we first reduce the terms with denominators $s_{134}^2s_{34}^2$ and $s_{234}^2s_{34}^2$.  The quark-emission term contains the trace
\begin{equation}
  \Tr\!\left[\slashed q\gamma^\alpha\slashed Q_q\gamma^\mu Z_-
  \slashed{\bar q}\gamma^\nu Z_-\slashed Q_q\gamma^\beta\right]
  \Tr\!\left[\slashed k\gamma_\alpha\slashed{\bar k}\gamma_\beta\right],
  \qquad Q_q=q+k+\bar k .
\end{equation}
After moving $Z_-$ through the massless spinor chain, the chiral-odd part drops out and the trace is proportional to $g_V^2+g_A^2$.  The contraction with the transverse polarization tensor can be performed in two steps.  The $-g_{\mu\nu}$ part gives
\begin{align}
  -g_{\mu\nu}
  \Tr\!\left[\slashed q\gamma^\alpha\slashed Q_q\gamma^\mu
  \slashed{\bar q}\gamma^\nu\slashed Q_q\gamma^\beta\right]
  &=2\Tr\!\left[\slashed q\gamma^\alpha\slashed Q_q
     \slashed{\bar q}\slashed Q_q\gamma^\beta\right] .
\end{align}
The part of Eq.~\eqref{eq:unresolved-trace} proportional to $-g_{\alpha\beta} k\cdot\bar k$ then yields
\begin{align}
  -8 k\cdot\bar k
  \Tr\!\left[\slashed q\gamma^\alpha\slashed Q_q
     \slashed{\bar q}\slashed Q_q\gamma_\alpha\right]
  &=16 k\cdot\bar k
  \Tr\!\left[\slashed q\slashed Q_q\slashed{\bar q}\slashed Q_q\right]
  \nn
  &=32 s_{34}(1+\frac{2 \mu^2}{s_{34}})
  \left[2(q\cdot Q_q)(\bar q\cdot Q_q)-q\cdot\bar q\,Q_q^2\right].
  \label{eq:uncrossed-trace-reduction}
\end{align}
Using
\begin{equation}
  2q\cdot Q_q=s_{134}-s_{34},
  \qquad
  2\bar q\cdot Q_q=s-s_{134},
  \qquad
  Q_q^2=s_{134},
\end{equation}
Eq.~\eqref{eq:uncrossed-trace-reduction} becomes
\begin{equation}
  16s_{34}(1+\frac{2\mu^2}{s_{34}})\left(s_{134}s_{234}-s\,s_{34}\right).
\end{equation}
Including the denominator $s_{134}^2s_{34}^2$ gives the compact integrand
\begin{equation}
  \mathcal{I}_{qq}
  =16(g_V^2+g_A^2)(1+\frac{2\mu^2}{s_{34}})
  \frac{s_{134}s_{234}-s\,s_{34}}{s_{34}s_{134}^2}
  =16(g_V^2+g_A^2)s(1+\frac{2\mu^2}{s_{34}})
  \left[-\frac{1}{s_{134}^2}
  +\frac{s_{234}}{s\,s_{34}s_{134}}\right].
  \label{eq:Iqq}
\end{equation}
The antiquark-emission contribution is obtained by $s_{134}\leftrightarrow s_{234}$,
\begin{equation}
  \mathcal{I}_{\bar q\bar q}
  =16(g_V^2+g_A^2)s(1+\frac{2\mu^2}{s_{34}})
  \left[-\frac{1}{s_{234}^2}
  +\frac{s_{134}}{s\,s_{34}s_{234}}\right].
  \label{eq:Iqbqb}
\end{equation}
Regarding the interference contribution, it contains the denominator $s_{134}s_{234}s_{34}^2$.  The trace structure is
\begin{equation}
  \Tr\!\left[\slashed q\gamma^\beta(\slashed p-\slashed{\bar q})(\slashed p-\slashed q)\gamma^\alpha\slashed{\bar q}\right]
  \Tr\!\left[\gamma_\beta\slashed k\gamma_\alpha\slashed{\bar k}\right].
\end{equation}
after contraction with the $g_{\mu\nu}$ part of the polarization tensor.  The contribution of the  $g_{\alpha\beta}k\cdot\bar k$ component of the unresolved-pair trace gives
\begin{equation}
  16(q\cdot\bar q)(p-q)\cdot(p-\bar q)(k\cdot\bar k)=2s_{12}(s_{134}+s_{234}+s_{12})s_{34}(1+\frac{2\mu^2}{s_{34}}).
\end{equation}
In this case, the interference integrand gives 
\begin{align}
  \mathcal{I}_{q\bar q}
  &=32(g_V^2+g_A^2) (1+\frac{2\mu^2}{s_{34}})s
  \left[
  \frac{s}{s_{34}s_{134}s_{234}}
  +\frac{2}{s_{134}s_{234}}
  +\frac{s_{34}}{s\,s_{134}s_{234}}
  \right.
  \nn
  &\hspace{3.2cm}\left.
  -\frac{1}{s_{34}s_{134}}
  -\frac{1}{s_{34}s_{234}}
  -\frac{1}{s\,s_{134}}
  -\frac{1}{s\,s_{234}}
  \right].
  \label{eq:Iinterference}
\end{align}
The remaining contributions proportional to
$k_\alpha\bar k_\beta$ and $k_\beta\bar k_\alpha$ can be accounted for using the Ward identity of the process under consideration. To this end, we define $K=k+\bar k$ and factorize the phase space as
\begin{equation}
 d\Phi_4=\frac{dK^2}{2\pi}\,
 d\Phi_3(p;q,\bar q,K)\,d\Phi_2(K;k,\bar k).
 \label{eq:real-phase-space-factorization}
\end{equation}
where $d\Phi_3$ relates to the part of the process involving the $Z q{\bar q}$ vertex with gluon emission from either fermion leg  and $d\Phi_2$ to the unresolved soft emission pair. At fixed $K$, the angular average of terms proportional to $k_\alpha\bar k_\beta$ and $k_\beta\bar k_\alpha$ is given by
\begin{align}
 \left\langle
 k_\alpha\bar k_\beta+k_\beta\bar k_\alpha
 \right\rangle_{\Omega_K}
 &=\frac13K_\alpha K_\beta+\frac16K^2g_{\alpha\beta}.
 \end{align}
 Therefore, the average of the full contribution (eq.~\ref{eq:unresolved-trace}) is 
 \begin{align}
 \left\langle L_{\alpha\beta}\right\rangle_{\Omega_K}
 &=\frac43\left(K_\alpha K_\beta-K^2g_{\alpha\beta}\right).
 \label{eq:pair-angular-average}
\end{align}
At this stage we can recall that the Ward identity applies to the coherent sum of the quark- and antiquark-emission currents \cite{Catani:1999ss}. If $H^{\alpha\beta}$ denotes the corresponding tensor related to the hard part of the process, then $K_\alpha H^{\alpha\beta}=K_\beta H^{\alpha\beta}=0$.
Consequently,
\begin{equation}
 H^{\alpha\beta}\left\langle L_{\alpha\beta}\right\rangle_{\Omega_K}
 =\frac23\,
H^{\alpha\beta}\left(-2K^2g_{\alpha\beta}\right),
 \label{eq:two-thirds-angular}
\end{equation}
where we explicitly wrote $-2K^2g_{\alpha\beta}$ which is the metric-only contribution $-4g_{\alpha\beta}k\cdot\bar k$.  Therefore, we see that we can account for the contributions stemming from $k_\alpha\bar k_\beta+k_\beta\bar k_\alpha$ as a global $2/3$ factor in the end of the calculation, in which only the the metric proportional piece was kept. 

Finally, we need to evaluate the phase-space integral. We define the four-particle phase-space measure by
\begin{equation}
  \PS(p;k_1,k_2,k_3,k_4)
  =(2\pi)^4\delta^{(4)}\!\left(p-\sum_{i=1}^4k_i\right)
  \prod_{i=1}^4\frac{\dd^3\mathbf{k}_i}{(2\pi)^3 2E_i} .
\end{equation}
It is convenient to define the regulator weight
\begin{equation}
w(\mu^2)\equiv 1+\frac{2\mu^2}{s_{34}},
\label{eq:real-weight}
\end{equation}
and to express the result in terms of the following phase-space integrals:
\begin{align}
  R_1&=\int\PS\,w(\mu^2)\,\frac12\left(\frac{1}{s\,s_{134}}+\frac{1}{s\,s_{234}}\right),
  &
  R_2&=\int\PS\,w(\mu^2)\,\frac{1}{s_{134}s_{234}},
  \nn
  R_3&=\int\PS\,w(\mu^2)\,\frac{s_{34}}{s\,s_{134}s_{234}},
  &
  R_4&=\int\PS\,w(\mu^2)\,\frac12\left(\frac{1}{s_{134}^2}+\frac{1}{s_{234}^2}\right),
  \nn
  R_5&=\int\PS\,w(\mu^2)\,\frac12\left(\frac{1}{s_{34}s_{134}}+\frac{1}{s_{34}s_{234}}\right),
  &
  R_6&=\int\PS\,w(\mu^2)\,\frac12\left(\frac{s_{234}}{s\,s_{34}s_{134}}+
\frac{s_{134}}{s\,s_{34}s_{234}}\right),
  \nn
  R_7&=\int\PS\,w(\mu^2)\,\frac{s}{s_{34}s_{134}s_{234}},
  \nn
  R_8&=\int\PS\,w(\mu^2)\,\frac{1}{s_{34}},
  \label{eq:Rbasis}
\end{align}

Using the definitions in Eq.~\eqref{eq:Rbasis}, the sum of the two uncrossed terms is
\begin{equation}
  \int\PS\left(\mathcal{I}_{qq}+\mathcal{I}_{\bar q\bar q}\right)
  =32(g_V^2+g_A^2)s\left(R_6-R_4\right),
  \label{eq:uncrossed-R}
\end{equation}
while the interference contribution is
\begin{equation}
  \int\PS\,\mathcal{I}_{q\bar q}
  =32(g_V^2+g_A^2)s
  \left(R_7+2R_2+R_3-2R_5-2R_1\right).
  \label{eq:interference-R}
\end{equation}
Therefore the real-emission contribution to the vector-current projection is
\begin{equation}
  \int\PS\,|\M_q+\M_{\bar q}|^2
  =\frac{64}{3}(g_V^2+g_A^2)s
  \left(-2R_1+2R_2+R_3-R_4-2R_5+R_6+R_7\right)
  \label{eq:real-final-R}
\end{equation}
up to the common factor $\mathcal{N}_{\rm real}$. The global factor of $2/3$ obtained in Eq.\ref{eq:two-thirds-angular} is already included in the result.

The dimensionless normalized counterparts of the phase-space integrals introduced above are the integrals used in Ref.~\cite{Page:2018ljf}. We have checked numerically the analytical results reported there and reproduce the corresponding formulas in Appendix~\ref{app:real} for convenience. The two phase-space conventions are related by
\begin{equation}
 d\Phi_4=(2\pi)^{-8}d\widetilde\Phi_4,
 \qquad
 R_i=\frac{\pi^3}{(2\pi)^8}\widetilde R_i
     =\frac{1}{256\pi^5}\widetilde R_i.
 \label{eq:R-convention-conversion}
\end{equation}

\subsection{\texorpdfstring{$Z\rightarrow q\bar q$}{Z to q qbar} decay width at NNLO}

The $Z\to q\bar q$ decay width through NNLO in the strong coupling, restricted to the large-$N_f$ contribution considered here, is assembled in two steps. First, one evaluates the two-particle phase-space integral of the coherent sum of virtual amplitudes through order $e^2g_s^4$, retaining the terms proportional to $N_f$. Second, one evaluates the real-emission phase-space integrals at the same perturbative order. The real sector contains the NNLO four-quark contribution derived in the previous subsection and, through the re-expansion of the NLO result, the contribution induced by the running of the strong coupling.

The NLO decay width, including the virtual correction and the one-gluon real-emission channel $q\bar q g$, was obtained within IReg in Ref.~\cite{Rosado:2023ist} and reads
\begin{equation}
\label{gamNLO}
\Gamma^{\rm NLO}=\Gamma_{\rm tree}\left(1+\frac{3C_F\alpha_s}{4\pi}\right),
\end{equation}
with
\begin{equation}
\Gamma_{\rm tree}=\frac{e^2(g_V^2+g_A^2)m_ZN_{c}}{12\pi\sin^2(2\omega)}.
\label{eq:Gamma-tree}
\end{equation}
In the large-$N_f$ limit the gluonic contribution to the running is omitted, so that the coupling evolves in a QED-like form. Expanding to first order in $N_f$ gives
\begin{align}
\alpha_s(\mu^2)
&=\frac{\alpha_s(\mu'^2)}{1-\dfrac{N_f}{6\pi}\alpha_s(\mu'^2)\ln\!\left(\dfrac{\mu^2}{\mu'^2}\right)}
\nonumber\\
&\simeq \alpha_s(m_Z^2)\left[1+\frac{N_f}{6\pi}\alpha_s(m_Z^2)\ln\!\left(\frac{\mu^2}{m_Z^2}\right)\right],
\label{coev}
\end{align}
where in the second line the reference scale has been chosen as $\mu'^2=m_Z^2=s$. The resulting $N_f$-dependent term induced from the NLO contribution is
\begin{equation}
\Gamma_{\rm ind}=\Gamma_{\rm tree}\,
2C_FN_f\frac{\alpha_s^2(m_Z^2)}{(4\pi)^2}L,
\qquad
L=\ln\!\left(\frac{\mu^2}{s}\right).
\label{eq:Gamma-induced}
\end{equation}
The auxiliary mass $\mu$ appearing in $L$ is the common mass parameter used to expose the real--virtual infrared cancellation; it is distinct from the IReg UV subtraction scale $\lambda$ introduced in Sec.~2.4.

With this organization, the virtual contribution is obtained from
\begin{align}
\int d\Phi_2\sum_{\rm spins}\left|\mathcal M_Z^{(0)}+V_Z^{(1)}+V_Z^{(2)}\right|^2
={}&\int d\Phi_2\sum_{\rm spins}\Big[
|\mathcal M_Z^{(0)}|^2
+2\Ree\!\left(\mathcal M_Z^{(0)\,*}V_Z^{(1)}\right)
\nonumber\\
&\hspace{2.6cm}
+2\Ree\!\left(\mathcal M_Z^{(0)\,*}V_Z^{(2)}\right)
\Big]_{N_f},
\label{vir}
\end{align}
where
\begin{equation}
d\Phi_2(p;q,\bar q)=(2\pi)^4\delta^{(4)}(p-q-\bar q)
\frac{d^3\mathbf q}{(2\pi)^3 2E_q}
\frac{d^3\mathbf{\bar q}}{(2\pi)^3 2E_{\bar q}}.
\label{eq:two-body-ps}
\end{equation}
The Born term together with the one-loop term in Eq.~\eqref{vir} forms the virtual part of the NLO correction, which combines with the one-gluon real-emission contribution to give Eq.~\eqref{gamNLO}. The term $|V_Z^{(1)}|^2$ is of order $e^2g_s^4$, but it carries the $C_F^2$ color structure and no factor of $N_f$; it is therefore outside the $C_FT_FN_f$ contribution isolated in the present calculation.

The two-loop NNLO contribution is obtained from the interference involving $V_Z^{(2)}$. The required scalar integrals were defined in Eq.~\eqref{eq:Aq}; they were evaluated numerically and compared with the analytic expressions first obtained in Ref.~\cite{Page:2018ljf}. The resulting virtual contribution is
\begin{equation}
\Gamma_V^{(2)}
=\Gamma_{\rm tree}\frac{4}{9}C_FN_f
\frac{\alpha_s^2(m_Z^2)}{(4\pi)^2}
\Ree\!\left[
L'^3+\frac{19}{2}L'^2
+L'\left(\frac{265}{6}+\pi^2\right)
-12\zeta_3+\frac{19}{6}\pi^2+\frac{3355}{36}
\right],
\label{eq:Gamma-virtual-NNLO}
\end{equation}
with
\begin{equation}
L'=\ln\!\left(\frac{\mu^2}{-s-i0^+}\right).
\end{equation}

For the real correction it is convenient to use the dimensionless phase-space integrals $\widetilde R_i$ listed in Appendix~\ref{app:real}. The decay rate is obtained from the polarization-summed matrix element as
\begin{equation}
 \Gamma_R^{(2)}
 =\frac{1}{2\sqrt{s}}\,\frac13
 \int d\Phi_4
 \left|\mathcal M_q+\mathcal M_{\bar q}\right|^2,
 \qquad \sqrt{s}=m_Z,
 \label{eq:real-width-definition}
\end{equation}
where the factor $1/3$ is the average over the three polarizations of the Z-boson. Combining Eqs.~\eqref{eq:real-final-R},
\eqref{eq:R-convention-conversion}, and \eqref{eq:Gamma-tree}, gives
\begin{align}
  \Gamma_R^{(2)}
 =\frac{e^2(g_V^2+g_A^2)m_Z}{\sin^2(2\omega)}
 N_cC_FT_FN_f\,
 \frac{\alpha_s^2}{(4\pi)^2}\,
 \frac{32}{9\pi}\,\mathcal R,
 \label{eq:real-width-absolute}
\end{align}
where
$\mathcal R=\widetilde R_7+\widetilde R_6-2\widetilde R_5-
\widetilde R_4+\widetilde R_3+2\widetilde R_2-2\widetilde R_1$.

We finally obtain for the NNLO real contribution 
\begin{align}
\Gamma_R^{(2)}
={}&\Gamma_{\rm tree}\frac{\alpha_s^2(m_Z^2)}{(4\pi)^2}
\frac{64}{3}C_FN_f
\left(\widetilde R_7+\widetilde R_6-2\widetilde R_5-\widetilde R_4
+\widetilde R_3+2\widetilde R_2-2\widetilde R_1\right)
\nonumber\\
={}&\Gamma_{\rm tree}\frac{\alpha_s^2(m_Z^2)}{(4\pi)^2}
\frac{4}{9}C_FN_f\left[
-L^3-\frac{19}{2}L^2
-L\left(\frac{146}{3}-2\pi^2\right)
+30\zeta_3+\frac{19}{3}\pi^2-\frac{2123}{18}
\right].
\label{eq:Gamma-real-NNLO}
\end{align}

Adding the induced, virtual, and real NNLO terms, all powers of the infrared logarithm cancel. The $N_f$-dependent correction to the tree-level result is therefore
\begin{equation}
\Gamma_{\rm tree}+\Gamma_{\rm ind}+\Gamma_V^{(2)}+\Gamma_R^{(2)}
=\Gamma_{\rm tree}\left[
1+\frac{\alpha_s^2(m_Z^2)}{(4\pi)^2}
C_FN_f\left(8\zeta_3-11\right)
\right],
\label{eq:final-width}
\end{equation}
in agreement with the corresponding large-$N_f$ result of Ref.~\cite{Chetyrkin:1996ela,Gehrmann-DeRidder:2004ttg,Page:2018ljf}.

\section{Summary and conclusions}

In this work we have further developed IReg as a viable framework for handling the UV and IR divergences that arise in multiloop decay and scattering processes in quantum field theory. Within the scope of the present investigation, we have applied IReg to the decay of the chiral gauge boson $Z$ into quarks and antiquarks through fourth order in the strong coupling $g_s$, corresponding to an NNLO calculation in the large-$N_f$ limit.

The calculation follows the characteristic IReg strategy of separating UV divergences locally, at the integrand level, before evaluating the UV-finite parts. The resulting divergent structures are treated by ordinary renormalization in a manner consistent with the BPHZ program \cite{AriasPerdomoEtAl2021,CherchigliaSampaioNemes2011}. In particular, the quark vacuum-polarization subgraph is isolated with its transverse structure intact, its BDI is removed by the corresponding lower-order counterterm, and the remaining outer-loop contribution is treated only after the subdivergence has been subtracted.

A second ingredient is the treatment of the $\gamma_5$ matrix in divergent integrals involving an open fermionic string. We employ a fixed right- or left-most ordering prescription in the open string. In the present process this is compatible with the more general IReg treatment of $\gamma_5$ based on the quasi-four-dimensional space $Q4S=4S\oplus X$. This construction provides the algebraic conditions required by the Quantum Action Principle and allows chiral structures to be handled unambiguously, as discussed in Refs.~\cite{Bruque:2018bmy,Cherchiglia:2021uce,Rosado:2024pyy,Cherchiglia:2026lps}. In processes in which $\gamma_5$ does not occur inside fermion propagators, the algebra generally simplifies sufficiently for right-most-position prescriptions or symmetrization of closed traces to be used \cite{Viglioni:2016nqc,Rosado:2023ist}.

A further step consists of parametrizing IR divergences with a fictitious mass in the propagators and in the corresponding cut phase space. This mass is taken to zero only after the virtual and real contributions have been combined. The regulator dependence therefore appears as logarithms and, where appropriate, finite terms that survive the vanishing-mass limit. The construction is organized so as to preserve gauge symmetry and the real--virtual correspondence implied by unitarity. In the present calculation this is seen explicitly in the cancellation of the infrared logarithms in the inclusive decay width, in accordance with the KLN theorem.

Underlying all of these operations is the IReg normal form discussed in Sec.~2.1. Numerator--denominator consistency, a fixed treatment of the Dirac algebra, and the systematic identification of surface terms are essential at two loops and beyond. At these orders one must simultaneously handle UV subdivergences and the overall divergence of the Feynman amplitude, while at the infrared end the singular contributions must cancel only after the corresponding phase-space integrals have been combined.

For the large-$N_f$ contribution studied here, the virtual fermion bubble and its real quark-pair cut provide a direct test of these requirements. After the fermion-bubble subdivergence is subtracted, the remaining $N_f$ contribution contains no additional UV divergence. The same fictitious mass can then be used consistently in the virtual and real sectors. Combining the induced, virtual, and real NNLO terms cancels all regulator-dependent logarithms and yields the coefficient $C_FN_f(8\zeta_3-11)$ multiplying $\alpha_s^2/(4\pi)^2$, in agreement with the known result. This agreement provides a non-trivial NNLO check of IReg for a chiral neutral-current process while keeping the loop algebra and momentum integrations in the physical dimension.

\appendix

\section{Wave-function renormalization in the large-\texorpdfstring{$N_f$}{Nf} limit at two loops}
\label{app:wfr}

Here we consider the diagrams in Figs.~\ref{fig.c} and \ref{fig.d}, which determine the external-leg contribution. We insert the finite part of the gluon polarization tensor in Eq.~\eqref{eq:PiGI}; its BDI part is omitted because it is removed by the counterterm represented by Fig.~\ref{fig.d}, in accordance with the BPHZ renormalization program.

\begin{eqnarray}
    \label{wfr}
&& V^{\mu}_{\rm wfr}= \int_{k,x} \frac{\overline{u}(q) \gamma^\beta (\slashed{q}-\slashed{k}) \gamma^\alpha \Pi_{\alpha,\beta}^{f} \slashed{q} \gamma^\mu v(\overline{q})}{k^4 q^2 (q-k)^2}\nonumber \\
&&=\int_{k,x} {\overline{u}(q)\frac{(-q^2 +2k\cdot q -\slashed{q}\slashed{k}) \gamma^\mu}{q^2 (q-k)^2} \frac{F(x)}{D_0}} v(\overline{q}) -\int_{k,x}{\overline{u}(q)\frac{ \slashed{k}\slashed{q} \gamma^\mu}{q^2 k^2} \frac{F(x)}{2 D_0}} v(\overline{q})\nonumber \\
\end{eqnarray} 
where $\Pi^f$ stands for the tensor polarization without the $I_{\log}$.
Here the last integral is odd in the integration over $k$ and vanishes. The term going as $\slashed{q} \slashed{k}$ also vanishes upon using Dirac's equation of motion for massless quarks. Notice that although there is $q^2$ in the denominator, the Dirac algebra and $N/D$ consistency are executed first and finally a regulating mass assigned to the propagators. 

Regarding renormalization, one is interested in the $UV$ behavior and thus one separates according to IReg the BDI from finite terms in UV, yielding
\begin{equation}
V^{\mu}_{\rm wfr}= -\int_{k,x} \frac{\bar u(q)\gamma^\mu F(x)}{k^2D_0}\left(1-\frac{4(k\cdot q)^2}{q^2k^2}\right)v(\bar q)+\text{UV-finite terms}.
\label{wfrd}
\end{equation}
The UV content vanishes after the same tensor-BDI reduction and surface-term constraints used in the virtual-vertex calculation.

\section{Quark self-energy in the large-\texorpdfstring{$N_f$}{Nf} limit at two loops}
\label{app:self}

The quark self-energy entering the external-leg analysis of Figs.~\ref{fig.c} and \ref{fig.d} reads

\begin{eqnarray}
\label{self}
&&-i \Sigma(q)=\int_{k,x} \frac{\gamma^\beta (\slashed{q} -\slashed{k}) \gamma^\alpha \Pi_{\alpha,\beta}^f}{k^4 (q-k)^2}\nonumber \\
&&=-\frac{1}{2}(-\slashed{q} +3 \slashed{k} -2\frac{k\cdot q \slashed{k}}{k^2})\frac{1}{(q-k)^2 D_0} F(x) \nonumber \\
\end{eqnarray}
upon which one has again to separate BDI from finite integral contributions leading to the following combination 
\begin{eqnarray}
&&-i\Sigma(q)=-\frac{1}{2}\int_{k,x}(-\frac{\slashed{q}}{k^2 D_0} +4 \frac{k\cdot q \slashed{k}}{k^4 D_0}) F(x)+\mbox{finite}\\
\end{eqnarray}
and leads to a vanishing combination of BDI.

\section{The real integrals}
\label{app:real}
\newcommand\bqa{\begin{eqnarray}}
\newcommand\eqa{\end{eqnarray}}
\newcommand\nl{\nonumber \\}
\newcommand\qt{q_2}
\newcommand\qot{q_{1\!2}}
\newcommand\qots{q^2_{1\!2}}
\newcommand\qotsb{\bar q^2_{1\!2}}

In this appendix we provide a list of the real integrals needed to obtain the NNLO corrections discussed in this work. We follow the notation of \cite{Page:2018ljf}, which define
\bqa
&&{\tilde R_1} := \frac{1}{s \pi^3} \int {d^4 \tilde \Phi_4}
\,w(\mu^2)\frac{1}{s_{134}},  \nl
&&{\tilde R_2} := \frac{1}{\pi^3} \int {d^4 \tilde \Phi_4}
\,w(\mu^2)\frac{1}{s_{134} s_{234}},  \nl
&&{\tilde R_3} :=\frac{1}{s \pi^3}   \int {d^4 \tilde \Phi_4}
\,w(\mu^2)\frac{s_{34}}{s_{134} s_{234}},  \nl
&&{\tilde R_4} := \frac{1}{\pi^3} \int {d^4 \tilde \Phi_4}
\,w(\mu^2)\frac{1}{s^2_{134}},  \nl
&&{\tilde R_5} := \frac{1}{\pi^3} \int {d^4 \tilde \Phi_4}
\,w(\mu^2)\frac{1}{s_{134} s_{34}}, \nl
&&{\tilde R_6} := \frac{1}{s \pi^3}  \int {d^4 \tilde \Phi_4}
\,w(\mu^2)\frac{s_{234}}{s_{134} s_{34}}, \nl
&&{\tilde R_7} := \frac{s}{\pi^3}  \int {d^4 \tilde \Phi_4}
\,w(\mu^2)\frac{1}{s_{34} s_{134} s_{234}}, \nl
&&{\tilde R_8} := \frac{1}{s \pi^3} \int {d^4 \tilde \Phi_4}
\,w(\mu^2)\frac{1}{s_{34}},
\eqa
with $w(\mu^2)=1+\dfrac{2\mu^2}{s_{34}}$, consistently with Eq.~\eqref{eq:real-weight}.

To compute the ${\tilde R}_i$s we employ the phase-space parametrization
\bqa
\label{eq:pspar}
\int d \tilde \Phi_4= \frac{s^2\pi^3}{8}
\int _{4 \epsilon^2}^{(1-2\epsilon)^2} \!\!\!\!\!\!dz\, \sqrt{1-4 \epsilon^2/z}
\int_z^1 dy
\int _{\frac{z}{y}}^{1-y+z} dx ,
\eqa
where
$
z = {s_{34}}/{s}
$,
$
y = {s_{234}}/{s}
$,
$
x = {s_{134}}/{s}
$,
and 
$
\epsilon^2 = {\mu^2}/{s}
$. The results were first presented in \cite{Page:2018ljf}, which we checked numerically. The analytic formulas, in the limit $\mu^{2}\rightarrow 0$, are given by
\bqa
    {\tilde R_1}&=& \frac{1}{32},\nl
    {\tilde R_2}&=& \frac{\pi^2}{48} -\frac{1}{8},\nl
    {\tilde R_3}&=& -\frac{\pi^2}{96}+\frac{7}{64},\nl
    {\tilde R_4}&=& -\frac{L}{16}-\frac{25}{96} ,\nl
    {\tilde R_5}&=&  \frac{L^2}{16}+\frac{11}{24}L
                     -\frac{\pi^2}{48}+\frac{85}{72},\nl
    {\tilde R_6}&=&  \frac{L^2}{32}+\frac{11}{48}L
                     -\frac{\pi^2}{96}+\frac{349}{576} ,\nl
    {\tilde R_7}&=&  -\frac{L^3}{48}-\frac{5}{48}L^2
    +L \left(\frac{\pi^2}{24}-\frac{7}{18}\right)
    +\frac{5}{8} \zeta_3+\frac{5}{72}\pi^2-\frac{41}{54},\nl
    {\tilde R_8}&=& {\tilde R_4},
\eqa
with $L=\ln(\mu^{2}/s)$.

\section*{Acknowledgments}

R.J.C.R. and B.H. acknowlege support from Fundação para a Ci\^encia e Tecnologia (FCT) by project
10.54499/UIDB/04564/2020 (https://sciproj.ptcris.pt/157582UID), and by project
10.54499/UIDP/04564/2020 (https://sciproj.ptcris.pt/157889UID).  R.J.C.R. acknowledges support by grant FCT 2020.07172.BD and  IT137-24-350. A.C. acknowledges support from the National Council for Scientific and Technological Development – CNPq grant 446121\slash2024-0.  

\bibliographystyle{unsrt} 
\bibliography{references}

@article{Kinoshita1962,
  author  = {Kinoshita, Toichiro},
  title   = {Mass Singularities of Feynman Amplitudes},
  journal = {J. Math. Phys.},
  volume  = {3},
  pages   = {650--677},
  year    = {1962},
  doi     = {10.1063/1.1724268}
}

@article{LeeNauenberg1964,
  author  = {Lee, T. D. and Nauenberg, M.},
  title   = {Degenerate Systems and Mass Singularities},
  journal = {Phys. Rev.},
  volume  = {133},
  pages   = {B1549--B1562},
  year    = {1964},
  doi     = {10.1103/PhysRev.133.B1549}
}

@article{tHooftVeltman1972,
  author  = {{'t Hooft}, Gerard and Veltman, Martinus},
  title   = {Regularization and Renormalization of Gauge Fields},
  journal = {Nucl. Phys. B},
  volume  = {44},
  pages   = {189--213},
  year    = {1972},
  doi     = {10.1016/0550-3213(72)90279-9}
}

@article{Catani1998,
  author        = {Catani, Stefano},
  title         = {The Singular Behaviour of QCD Amplitudes at Two-Loop Order},
  journal       = {Phys. Lett. B},
  volume        = {427},
  pages         = {161--171},
  year          = {1998},
  doi           = {10.1016/S0370-2693(98)00332-3},
  eprint        = {hep-ph/9802439},
  archivePrefix = {arXiv}
}

@article{BattistelMotaNemes1998,
  author  = {Battistel, O. A. and Mota, A. L. and Nemes, M. C.},
  title   = {Consistency Conditions for Four-Dimensional Regularizations},
  journal = {Mod. Phys. Lett. A},
  volume  = {13},
  pages   = {1597--1610},
  year    = {1998},
  doi     = {10.1142/S0217732398001686}
}

@article{SampaioEtAl2006,
  author        = {Sampaio, M. D. and Scarpelli, A. P. Baeta
                   and Ottoni, J. E. and Nemes, M. C.},
  title         = {Implicit Regularization and Renormalization of QCD},
  journal       = {Int. J. Theor. Phys.},
  volume        = {45},
  pages         = {436--457},
  year          = {2006},
  doi           = {10.1007/s10773-006-9045-z},
  eprint        = {hep-th/0509102},
  archivePrefix = {arXiv}
}

@article{DiasEtAl2008,
  author        = {Dias, E. W. and Scarpelli, A. P. Baeta and Brito, L. C. T.
                   and Sampaio, M. and Nemes, M. C.},
  title         = {Implicit Regularization Beyond One Loop Order:
                   Gauge Field Theories},
  journal       = {Eur. Phys. J. C},
  volume        = {55},
  pages         = {667--681},
  year          = {2008},
  doi           = {10.1140/epjc/s10052-008-0614-6},
  eprint        = {0801.2703},
  archivePrefix = {arXiv},
  primaryClass  = {hep-th}
}

@article{CherchigliaSampaioNemes2011,
  author        = {Cherchiglia, A. L. and Sampaio, M. and Nemes, M. C.},
  title         = {Systematic Implementation of Implicit Regularization
                   for Multi-Loop Feynman Diagrams},
  journal       = {Int. J. Mod. Phys. A},
  volume        = {26},
  pages         = {2591--2635},
  year          = {2011},
  doi           = {10.1142/S0217751X11053419},
  eprint        = {1008.1377},
  archivePrefix = {arXiv},
  primaryClass  = {hep-th}
}

@article{FerreiraEtAl2012,
  author        = {Ferreira, L. C. and Cherchiglia, A. L. and Hiller, B.
                   and Sampaio, M. and Nemes, M. C.},
  title         = {Momentum Routing Invariance in Feynman Diagrams
                   and Quantum Symmetry Breakings},
  journal       = {Phys. Rev. D},
  volume        = {86},
  pages         = {025016},
  year          = {2012},
  doi           = {10.1103/PhysRevD.86.025016},
  eprint        = {1110.6186},
  archivePrefix = {arXiv},
  primaryClass  = {hep-th}
}

@article{AriasPerdomoEtAl2021,
  author        = {Arias-Perdomo, D. C. and Cherchiglia, A.
                   and Hiller, B. and Sampaio, M.},
  title         = {A Brief Review of Implicit Regularization
                   and Its Connection with the BPHZ Theorem},
  journal       = {Symmetry},
  volume        = {13},
  number        = {6},
  pages         = {956},
  year          = {2021},
  doi           = {10.3390/sym13060956},
  eprint        = {2105.08085},
  archivePrefix = {arXiv},
  primaryClass  = {hep-th}
}

@article{Rosado:2023ist,
    author = "Rosado, Ricardo J. C. and Cherchiglia, Adriano and Sampaio, Marcos and Hiller, Brigitte",
    title = "{Infrared subtleties and chiral vertices at NLO: an implicit regularization analysis}",
    eprint = "2305.07129",
    archivePrefix = "arXiv",
    primaryClass = "hep-ph",
    doi = "10.1140/epjc/s10052-023-12052-2",
    journal = "Eur. Phys. J. C",
    volume = "83",
    number = "9",
    pages = "879",
    year = "2023"
}

@article{Rosado:2024pyy,
    author = "Rosado, Ricardo J. C. and Cherchiglia, Adriano and Sampaio, Marcos and Hiller, Brigitte",
    title = "{Addressing $\gamma _5$ in nondimensional regularizations: a case study on the Bumblebee model}",
    eprint = "2404.15551",
    archivePrefix = "arXiv",
    primaryClass = "hep-ph",
    doi = "10.1140/epjc/s10052-025-13756-3",
    journal = "Eur. Phys. J. C",
    volume = "85",
    number = "1",
    pages = "41",
    year = "2025"
}

@article{DineSapirstein1979,
  author  = {Dine, Michael and Sapirstein, Jonathan},
  title   = {Higher-Order Quantum-Chromodynamic Corrections
             in Electron-Positron Annihilation},
  journal = {Phys. Rev. Lett.},
  volume  = {43},
  pages   = {668--671},
  year    = {1979},
  doi     = {10.1103/PhysRevLett.43.668}
}

@article{Cherchiglia:2026lps,
  author  = {Cherchiglia, Adriano and Rosado, Ricardo J. C. and Sampaio, Marcos and Hiller, Brigitte},
  title   = {Chiral Symmetry in Implicit Regularization: A Review},
  journal = {Symmetry},
  volume  = {18},
  number  = {1},
  pages   = {160},
  year    = {2026},
  doi     = {10.3390/sym18010160}
}

@article{Jones:1982zf,
    author = "Jones, D. R. T. and Leveille, J. P.",
    title = "{Dimensional Regularization and the Two Loop Axial Anomaly in Abelian, Nonabelian and Supersymmetric Gauge Theories}",
    reportNumber = "UM-HE-81-74-REV, UM-HE-81-74",
    doi = "10.1016/0550-3213(83)90549-7",
    journal = "Nucl. Phys. B",
    volume = "206",
    pages = "473",
    year = "1982",
    note = "[Erratum: Nucl.Phys.B 222, 517 (1983)]"
}

@article{ZurbanoFernandez:2020cco,
    author = "Zurbano Fernandez, I. and others",
    editor = {B{\'e}jar Alonso, I. and Br{\"u}ning, O. and Fessia, P. and Rossi, L. and Tavian, L. and Zerlauth, M.},
    title = "{High-Luminosity Large Hadron Collider (HL-LHC): Technical design report}",
    reportNumber = "CERN-2020-010",
    doi = "10.23731/CYRM-2020-0010",
    volume = "10/2020",
    month = "12",
    year = "2020"
}

@article{FCC:2018evy,
    author = "Abada, A. and others",
    collaboration = "FCC",
    title = "{FCC-ee: The Lepton Collider}: {Future Circular Collider Conceptual Design Report Volume 2}",
    reportNumber = "CERN-ACC-2018-0057",
    doi = "10.1140/epjst/e2019-900045-4",
    journal = "Eur. Phys. J. ST",
    volume = "228",
    number = "2",
    pages = "261--623",
    year = "2019"
}

@article{vonManteuffel:2025swv,
    author = {von Manteuffel, Andreas and St{\"o}ckinger, Dominik and Wei{\ss}wange, Matthias},
    title = "{Four-Loop Renormalisation of Chiral Gauge Theories with Non-Anticommuting $\gamma_5$ in the BMHV Scheme}",
    eprint = "2506.12253",
    archivePrefix = "arXiv",
    primaryClass = "hep-ph",
    doi = "10.1007/JHEP08(2025)088",
    journal = "JHEP",
    volume = "08",
    pages = "088",
    year = "2025"
}

@article{Cornella:2022hkc,
    author = "Cornella, Claudia and Feruglio, Ferruccio and Vecchi, Luca",
    title = "{Gauge invariance and finite counterterms in chiral gauge theories}",
    eprint = "2205.10381",
    archivePrefix = "arXiv",
    primaryClass = "hep-ph",
    doi = "10.1007/JHEP02(2023)244",
    journal = "JHEP",
    volume = "02",
    pages = "244",
    year = "2023"
}

@article{Rios-Sanchez:2024xtv,
    author = "Rios-Sanchez, Jose and Sborlini, German",
    title = "{Toward multiloop local renormalization within causal loop-tree duality}",
    eprint = "2402.13995",
    archivePrefix = "arXiv",
    primaryClass = "hep-th",
    doi = "10.1103/PhysRevD.109.125004",
    journal = "Phys. Rev. D",
    volume = "109",
    number = "12",
    pages = "125004",
    year = "2024"
}

@article{Kuhler:2025znv,
    author = {K{\"u}hler, Paul and St{\"o}ckinger, Dominik},
    title = "{Two-loop renormalization of a chiral SU(2) gauge theory in dimensional regularization with non-anticommuting {\ensuremath{\gamma}}$_{5}$}",
    eprint = "2504.06080",
    archivePrefix = "arXiv",
    primaryClass = "hep-ph",
    doi = "10.1007/JHEP07(2025)082",
    journal = "JHEP",
    volume = "07",
    pages = "082",
    year = "2025"
}

@article{Ebert:2024xpy,
    author = {Ebert, Paul L. and K{\"u}hler, Paul and St{\"o}ckinger, Dominik and Wei{\ss}wange, Matthias},
    title = "{Shedding light on evanescent shadows {\textemdash} Exploration of non-anticommuting {\ensuremath{\gamma}}$_{5}$ in Dimensional Regularisation}",
    eprint = "2411.02543",
    archivePrefix = "arXiv",
    primaryClass = "hep-ph",
    doi = "10.1007/JHEP01(2025)114",
    journal = "JHEP",
    volume = "01",
    pages = "114",
    year = "2025"
}

@article{Stockinger:2023ndm,
    author = {St{\"o}ckinger, Dominik and Wei{\ss}wange, Matthias},
    title = "{Full three-loop renormalisation of an abelian chiral gauge theory with non-anticommuting $\gamma_{5}$ in the BMHV scheme}",
    eprint = "2312.11291",
    archivePrefix = "arXiv",
    primaryClass = "hep-ph",
    doi = "10.1007/JHEP02(2024)139",
    journal = "JHEP",
    volume = "02",
    pages = "139",
    year = "2024"
}

@article{Belusca-Maito:2023wah,
    author = {B{\'e}lusca-Ma{\"\i}to, Herm{\`e}s and Ilakovac, Amon and K{\"u}hler, Paul and Mador-Bozinovic, Marija and St{\"o}ckinger, Dominik and Wei{\ss}wange, Matthias},
    title = "{Introduction to Renormalization Theory and Chiral Gauge Theories in Dimensional Regularization with Non-Anticommuting {\ensuremath{\gamma}}$_{5}$}",
    eprint = "2303.09120",
    archivePrefix = "arXiv",
    primaryClass = "hep-ph",
    doi = "10.3390/sym15030622",
    journal = "Symmetry",
    volume = "15",
    number = "3",
    pages = "622",
    year = "2023"
}

@article{Cherchiglia:2012zp,
    author = "Cherchiglia, A. L. and Cabral, L. A. and Nemes, M. C. and Sampaio, Marcos",
    title = "{(Un)determined finite regularization dependent quantum corrections: the Higgs boson decay into two photons and the two photon scattering examples}",
    eprint = "1210.6164",
    archivePrefix = "arXiv",
    primaryClass = "hep-th",
    doi = "10.1103/PhysRevD.87.065011",
    journal = "Phys. Rev. D",
    volume = "87",
    number = "6",
    pages = "065011",
    year = "2013"
}

@article{Gnendiger:2017pys,
    author = "Gnendiger, C. and others",
    title = "{To ${d}$, or not to ${d}$: recent developments and comparisons of regularization schemes}",
    eprint = "1705.01827",
    archivePrefix = "arXiv",
    primaryClass = "hep-ph",
    reportNumber = "PSI-PR-17-06, ZU-TH-10-17, TUM-HEP-1081-17, IFIC-17-17, TIF-UNIMI-2017-4",
    doi = "10.1140/epjc/s10052-017-5023-2",
    journal = "Eur. Phys. J. C",
    volume = "77",
    number = "7",
    pages = "471",
    year = "2017"
}

@article{Pereira:2022mne,
    author = "Pereira, Ana and Cherchiglia, Adriano and Sampaio, Marcos and Hiller, Brigitte",
    title = "{Higgs boson decay into gluons in a 4D regularization: IR cancellation without evanescent fields to NLO}",
    eprint = "2207.05187",
    archivePrefix = "arXiv",
    primaryClass = "hep-ph",
    doi = "10.1140/epjc/s10052-023-11173-y",
    journal = "Eur. Phys. J. C",
    volume = "83",
    number = "1",
    pages = "73",
    year = "2023"
}

@article{TorresBobadilla:2020ekr,
    author = "Torres Bobadilla, W. J. and others",
    title = "{May the four be with you: Novel IR-subtraction methods to tackle NNLO calculations}",
    eprint = "2012.02567",
    archivePrefix = "arXiv",
    primaryClass = "hep-ph",
    reportNumber = "IFIC/20-54; MPP-2020-218; TTP20-043; P3H-20-079",
    doi = "10.1140/epjc/s10052-021-08996-y",
    journal = "Eur. Phys. J. C",
    volume = "81",
    number = "3",
    pages = "250",
    year = "2021"
}

@article{Cherchiglia:2015vaa,
    author = "Cherchiglia, A. L. and Sampaio, Marcos and Hiller, B. and Scarpelli, A. P. Ba{\^e}ta",
    title = "{Subtleties in the beta function calculation of N=1 supersymmetric gauge theories}",
    eprint = "1508.05421",
    archivePrefix = "arXiv",
    primaryClass = "hep-th",
    doi = "10.1140/epjc/s10052-015-3859-x",
    journal = "Eur. Phys. J. C",
    volume = "76",
    number = "2",
    pages = "47",
    year = "2016"
}

@article{Breitenlohner:1977hr,
    author = "Breitenlohner, P. and Maison, D.",
    title = "{Dimensional Renormalization and the Action Principle}",
    doi = "10.1007/BF01609069",
    journal = "Commun. Math. Phys.",
    volume = "52",
    pages = "11--38",
    year = "1977"
}

@article{Bruque:2018bmy,
    author = "Bruque, A. M. and Cherchiglia, A. L. and P{\'e}rez-Victoria, M.",
    title = "{Dimensional regularization vs methods in fixed dimension with and without $\gamma_5$}",
    eprint = "1803.09764",
    archivePrefix = "arXiv",
    primaryClass = "hep-ph",
    reportNumber = "CERN-TH-2018-058",
    doi = "10.1007/JHEP08(2018)109",
    journal = "JHEP",
    volume = "08",
    pages = "109",
    year = "2018"
}

@article{Cherchiglia:2020iug,
    author = "Cherchiglia, A. and Arias-Perdomo, D. C. and Vieira, A. R. and Sampaio, M. and Hiller, B.",
    title = "{Two-loop renormalisation of gauge theories in $4D$ Implicit Regularisation: transition rules to dimensional methods}",
    eprint = "2006.10951",
    archivePrefix = "arXiv",
    primaryClass = "hep-ph",
    doi = "10.1140/epjc/s10052-021-09259-6",
    journal = "Eur. Phys. J. C",
    volume = "81",
    number = "5",
    pages = "468",
    year = "2021"
}

@article{Cherchiglia:2021yxz,
    author = "Cherchiglia, Adriano",
    title = "{Two-loop gauge coupling $\beta$-function in a four-dimensional framework: the Standard Model case}",
    eprint = "2110.01739",
    archivePrefix = "arXiv",
    primaryClass = "hep-ph",
    doi = "10.21468/SciPostPhysProc.7.043",
    journal = "SciPost Phys. Proc.",
    volume = "7",
    pages = "043",
    year = "2022"
}

@article{Fargnoli:2010mf,
    author = "Fargnoli, H. G. and Hiller, B. and Scarpelli, A. P. Baeta and Sampaio, Marcos and Nemes, M. C.",
    title = "{Regularization Independent Analysis of the Origin of Two Loop Contributions to N=1 Super Yang-Mills Beta Function}",
    eprint = "1009.2976",
    archivePrefix = "arXiv",
    primaryClass = "hep-th",
    doi = "10.1140/epjc/s10052-011-1633-2",
    journal = "Eur. Phys. J. C",
    volume = "71",
    pages = "1633",
    year = "2011"
}

@article{Viglioni:2016nqc,
    author = "Viglioni, A. C. D. and Cherchiglia, A. L. and Vieira, A. R. and Hiller, Brigitte and Sampaio, Marcos",
    title = "{$\gamma_{5}$ algebra ambiguities in Feynman amplitudes: Momentum routing invariance and anomalies in $D=4$ and $D=2$}",
    eprint = "1606.01772",
    archivePrefix = "arXiv",
    primaryClass = "hep-th",
    doi = "10.1103/PhysRevD.94.065023",
    journal = "Phys. Rev. D",
    volume = "94",
    number = "6",
    pages = "065023",
    year = "2016"
}

@article{Heinemeyer:2021rgq,
    author = {Heinemeyer, Sven and Jadach, Stanislaw and Reuter, J{\"u}rgen},
    title = "{Theory requirements for SM Higgs and EW precision physics at the FCC-ee}",
    eprint = "2106.11802",
    archivePrefix = "arXiv",
    primaryClass = "hep-ph",
    reportNumber = "DESY-21-100",
    doi = "10.1140/epjp/s13360-021-01875-1",
    journal = "Eur. Phys. J. Plus",
    volume = "136",
    number = "9",
    pages = "911",
    year = "2021", 
    note = "[Erratum: Eur. Phys. J. Plus (2022)137:792]"
}

@article{stockinger2005regularization,
  title={Regularization by dimensional reduction: consistency, quantum action principle, and supersymmetry},
  author={St{\"o}ckinger, Dominik},
  journal={Journal of High Energy Physics},
  volume={2005},
  number={03},
  pages={076},
  year={2005},
  publisher={IOP Publishing}
}

@article{hernandez2016towards,
  title={Towards gauge theories in four dimensions},
  author={Hernandez-Pinto, Roger J and Sborlini, German FR and Rodrigo, German},
  journal={Journal of High Energy Physics},
  volume={2016},
  number={2},
  pages={1--14},
  year={2016},
  publisher={Springer}
}

@article{sborlini2016four,
  title={Four-dimensional unsubtraction from the loop-tree duality},
  author={Sborlini, German FR and Driencourt-Mangin, Felix and Hernandez-Pinto, Roger J and Rodrigo, German},
  journal={Journal of High Energy Physics},
  volume={2016},
  number={8},
  pages={1--42},
  year={2016},
  publisher={Springer}
}

@article{breitenlohner1977dimensionally,
  title={Dimensionally renormalized Green's functions for theories with massless particles. II},
  author={Breitenlohner, P and Maison, D},
  journal={Communications in Mathematical Physics},
  volume={52},
  number={1},
  pages={55--75},
  year={1977},
  publisher={Springer}
}

@article{korner1992practicable,
  title={A Practicable $\gamma$ 5-scheme in dimensional regularization},
  author={K{\"o}rner, J{\"u}rgen G and Kreimer, Dirk and Schilcher, Karl},
  journal={Zeitschrift f{\"u}r Physik C Particles and Fields},
  volume={54},
  pages={503--512},
  year={1992},
  publisher={Springer}
}

@article{kreimer1994role,
  title={The Role of $\gamma_5$ in Dimensional Regularization},
  author={Kreimer, Dirk},
  journal={arXiv preprint hep-ph/9401354},
  year={1994}
}

@article{larin1993renormalization,
  title={The Renormalization of the axial anomaly in dimensional regularization},
  author={Larin, SA},
  journal={Quarks-92, ed. D. Yu. Grigoriev, VA Matveev, VA Rubakov, PG Tinyakov, World Scientific},
  pages={201},
  year={1993},
  publisher={World Scientific}
}

@article{jegerlehner2001facts,
  title={Facts of life with $\gamma_5$},
  author={Jegerlehner, Fred},
  journal={The European Physical Journal C-Particles and Fields},
  volume={18},
  number={4},
  pages={673--679},
  year={2001},
  publisher={Springer}
}

@article{tsai2009advantage,
  title={The Advantage of Rightmost Ordering for gamma5 in Dimensional Regularization},
  author={Tsai, Er-Cheng},
  journal={arXiv preprint arXiv:0905.1479},
  year={2009}
}

@article{tsai2011gauge,
  title={Gauge Invariant Treatment of $\gamma$ 5 in the Scheme of’t Hooft and Veltman},
  author={Tsai, Er-Cheng},
  journal={Physical Review D},
  volume={83},
  number={2},
  pages={025020},
  year={2011},
  publisher={APS}
}

@article{belusca2021two,
  title={Two-loop application of the Breitenlohner-Maison/’t Hooft-Veltman scheme with non-anticommuting $\gamma$5: full renormalization and symmetry-restoring counterterms in an abelian chiral gauge theory},
  author={B{\'e}lusca-Ma{\"\i}to, Herm{\`e}s and Ilakovac, Amon and K{\"u}hler, Paul and Mador-Bo{\v{z}}inovi{\'c}, Marija and St{\"o}ckinger, Dominik},
  journal={Journal of High Energy Physics},
  volume={2021},
  number={11},
  pages={1--32},
  year={2021},
  publisher={Springer}
}

@article{ramirez2024rewording,
  title={Rewording theoretical predictions at colliders with vacuum amplitudes},
  author={Ram{\'\i}rez-Uribe, Selomit and Dhani, Prasanna K and Sborlini, German FR and Rodrigo, Germ{\'a}n},
  journal={Physical Review Letters},
  volume={133},
  number={21},
  pages={211901},
  year={2024},
  publisher={APS}
}

@article{signer2009using,
  title={Using dimensional reduction for hadronic collisions},
  author={Signer, Adrian and St{\"o}ckinger, Dominik},
  journal={Nuclear physics B},
  volume={808},
  number={1-2},
  pages={88--120},
  year={2009},
  publisher={Elsevier}
}

@article{Cherchiglia:2021uce,
    author = "Cherchiglia, Adriano",
    title = "{Step towards a consistent treatment of chiral theories at higher loop order: The abelian case}",
    eprint = "2106.14039",
    archivePrefix = "arXiv",
    primaryClass = "hep-ph",
    doi = "10.1016/j.nuclphysb.2023.116104",
    journal = "Nucl. Phys. B",
    volume = "987",
    pages = "116104",
    year = "2023"
}

@article{LTD:2024yrb,
    author = "Ram{\'\i}rez-Uribe, Selomit and Renter{\'\i}a-Olivo, Andr{\'e}s E. and Renter{\'\i}a-Estrada, David F. and de Lejarza, Jorge J. Mart{\'\i}nez and Dhani, Prasanna K. and Cieri, Leandro and Hern{\'a}ndez-Pinto, Roger J. and Sborlini, German F. R. and Torres Bobadilla, William J. and Rodrigo, Germ{\'a}n",
    collaboration = "LTD",
    title = "{Vacuum amplitudes and time-like causal unitary in the loop-tree duality}",
    eprint = "2404.05492",
    archivePrefix = "arXiv",
    primaryClass = "hep-ph",
    doi = "10.1007/JHEP01(2025)103",
    journal = "JHEP",
    volume = "01",
    pages = "103",
    year = "2025"
}

@article{Pittau:2012zd,
    author = "Pittau, Roberto",
    title = "{A four-dimensional approach to quantum field theories}",
    eprint = "1208.5457",
    archivePrefix = "arXiv",
    primaryClass = "hep-ph",
    doi = "10.1007/JHEP11(2012)151",
    journal = "JHEP",
    volume = "11",
    pages = "151",
    year = "2012"
}

@article{kilgore:2012tb,
  author        = "Kilgore, William B.",
  title         = "{The Four Dimensional Helicity Scheme Beyond One Loop}",
  eprint        = "1205.4015",
  archivePrefix = "arXiv",
  primaryClass  = "hep-ph",
  doi           = "10.1103/PhysRevD.86.014019",
  journal       = "Phys. Rev. D",
  volume        = "86",
  pages         = "014019",
  year          = "2012"
}

@article{Zimmermann:1969jw,
  author        = "Zimmermann, W.",
  title         = "{Convergence of Bogolyubov's method of renormalization in momentum space}",
  doi           = "10.1007/BF01645676",
  journal       = "Commun. Math. Phys.",
  volume        = "15",
  pages         = "208--234",
  year          = "1969"
}

@article{Bogoliubov:1957pq,
  author        = "Bogoliubov, N. N. and Parasiuk, O. S.",
  title         = "{On the multiplication of the causal function in the quantum theory of fields}",
  doi           = "10.1007/BF02392399",
  journal       = "Acta Math.",
  volume        = "97",
  pages         = "227--266",
  year          = "1957"
}

@article{Hepp:1966eg,
  author        = "Hepp, Klaus",
  title         = "{Proof of the Bogolyubov-Parasiuk theorem on renormalization}",
  doi           = "10.1007/BF01773358",
  journal       = "Commun. Math. Phys.",
  volume        = "2",
  pages         = "301--326",
  year          = "1966"
}

@article{Donati:2013voa,
    author = "Donati, Alice Maria and Pittau, Roberto",
    title = "{FDR, an easier way to NNLO calculations: a two-loop case study}",
    eprint = "1311.3551",
    archivePrefix = "arXiv",
    primaryClass = "hep-ph",
    doi = "10.1140/epjc/s10052-014-2864-9",
    journal = "Eur. Phys. J. C",
    volume = "74",
    pages = "2864",
    year = "2014"
}

@article{Page:2018ljf,
    author = "Page, Ben and Pittau, Roberto",
    title = "{NNLO final-state quark-pair corrections in four dimensions}",
    eprint = "1810.00234",
    archivePrefix = "arXiv",
    primaryClass = "hep-ph",
    doi = "10.1140/epjc/s10052-019-6865-6",
    journal = "Eur. Phys. J. C",
    volume = "79",
    number = "4",
    pages = "361",
    year = "2019"
}

@article{Pittau:2018frs,
    author = "Pittau, Roberto",
    title = "{Computing radiative corrections in four dimensions}",
    eprint = "1801.02337",
    archivePrefix = "arXiv",
    primaryClass = "hep-ph",
    doi = "10.5506/APhysPolBSupp.11.387",
    journal = "Acta Phys. Polon. Supp.",
    volume = "11",
    pages = "387",
    year = "2018"
}

@article{OlgosoRuiz:2024dzq,
    author = "Olgoso Ruiz, Pablo and Vecchi, Luca",
    title = "{Spurious gauge-invariance and {\ensuremath{\gamma}}$_{5}$ in dimensional regularization}",
    eprint = "2406.17013",
    archivePrefix = "arXiv",
    primaryClass = "hep-ph",
    doi = "10.1007/JHEP12(2024)080",
    journal = "JHEP",
    volume = "12",
    pages = "080",
    year = "2024"
}

@article{Catani:1999ss,
    author = "Catani, Stefano and Grazzini, Massimiliano",
    title = "{Infrared factorization of tree level QCD amplitudes at the next-to-next-to-leading order and beyond}",
    eprint = "hep-ph/9908523",
    archivePrefix = "arXiv",
    reportNumber = "CERN-TH-99-263, ETH-TH-99-22",
    doi = "10.1016/S0550-3213(99)00778-6",
    journal = "Nucl. Phys. B",
    volume = "570",
    pages = "287--325",
    year = "2000"
}

@article{Gehrmann-DeRidder:2004ttg,
    author = "Gehrmann-De Ridder, A. and Gehrmann, T. and Glover, E. W. Nigel",
    title = "{Infrared structure of e+ e- ---{\ensuremath{>}} 2 jets at NNLO}",
    eprint = "hep-ph/0403057",
    archivePrefix = "arXiv",
    reportNumber = "ZU-TH-03-04, IPPP-04-12, DCPT-04-24, NSF-KITP-04-32",
    doi = "10.1016/j.nuclphysb.2004.05.017",
    journal = "Nucl. Phys. B",
    volume = "691",
    pages = "195--222",
    year = "2004"
}

@article{Chetyrkin:1996ela,
    author = "Chetyrkin, K. G. and Kuhn, Johann H. and Kwiatkowski, A.",
    title = "{QCD corrections to the $e^{+} e^{-}$ cross-section and the $Z$ boson decay rate}",
    eprint = "hep-ph/9503396",
    archivePrefix = "arXiv",
    reportNumber = "LBL-36678-REV, TTP-94-32, LBL-36678, MPI-PHT-96-019",
    doi = "10.1016/S0370-1573(96)00012-9",
    journal = "Phys. Rept.",
    volume = "277",
    pages = "189--281",
    year = "1996"
}

\end{document}